\documentclass[
 reprint,
 superscriptaddress,
 footinbib,
 aps,
 amsmath,
 amssymb,
  longbibliography
]{revtex4-2}

\usepackage{graphicx} 
\usepackage{mathtools} 
\usepackage{physics}  
\usepackage{bbold} 
\usepackage{bm} 
\usepackage{xcolor, soul} 
\usepackage{hyperref}
\usepackage{ragged2e}
\usepackage{ulem}

\begin{document}

\title{Quantum theory of electrically levitated nanoparticle-ion systems: \\Motional dynamics and sympathetic cooling}
\author{Saurabh Gupta}
\affiliation{
 Institute for Theoretical Physics and Vienna Center for Quantum Science and Technology, Technical University of Vienna, Wiedner Hauptstraße 8-10, 1040 Vienna, Austria
}
\author{Bernard Faulend}
\affiliation{
 Institute for Theoretical Physics and Vienna Center for Quantum Science and Technology, Technical University of Vienna, Wiedner Hauptstraße 8-10, 1040 Vienna, Austria
}
\author{Dmitry S. Bykov}
\affiliation{Institut f\"ur Experimentalphysik, Universit\"at Innsbruck, Technikerstraße 25, 6020 Innsbruck, Austria}
\author{Tracy E. Northup}
\affiliation{Institut f\"ur Experimentalphysik, Universit\"at Innsbruck, Technikerstraße 25, 6020 Innsbruck, Austria}
\author{Carlos Gonzalez-Ballestero}
\email{carlos.gonzalez-ballestero@tuwien.ac.at}
\affiliation{
 Institute for Theoretical Physics and Vienna Center for Quantum Science and Technology, Technical University of Vienna, Wiedner Hauptstraße 8-10, 1040 Vienna, Austria
}

\date{\today}

\begin{abstract}
We develop the theory describing the quantum coupled dynamics of the center-of-mass motion of a nanoparticle and an ensemble of ions co-trapped in a dual-frequency linear Paul trap. We first derive analytical expressions for the motional frequencies and classical trajectories of both nanoparticle and ions. We then derive a quantum master equation for the ion-nanoparticle system and quantify the sympathetic cooling of the nanoparticle motion enabled by its Coulomb coupling to a continuously Doppler-cooled ion. We predict that motional cooling down to sub-kelvin temperatures is achievable in state-of-the-art experiments even in the absence of motional feedback and in the presence of micromotion. We then extend our analysis to an ensemble of $N$ ions, predicting a linear increase of the cooling rate as a function of $N$ and motional cooling of the nanoparticle down to tenths of millikelvin in current experimental platforms. Our work establishes the theoretical toolbox needed to explore the ion-assisted preparation of non-Gaussian motional states of levitated nanoparticles. 
\end{abstract}

\maketitle

\section{Introduction}\label{sec:introduction}

The preparation of macroscopic quantum states of the center-of-mass motion of nanoparticles levitated in high vacuum is a central goal in levitated optomechanics. Such states could shed light on the mechanisms behind the quantum-to-classical transition and enable tests of the quantum nature of gravity \cite{Millen_2020,CGB_2021}. Following demonstrations of ground-state motional cooling \cite{Delic_2020,Tebbenjohanns_2021,Magrini_2021,Kamba_2022,Ranfagni_2022,Piotrowski_2023,Dania_2025} and motional squeezing \cite{Rossi_2024,Kamba_2025} of an optically trapped nanoparticle, the next key milestone is the preparation of non-Gaussian motional states. However, for optically trapped nanoparticles, motional decoherence induced by laser recoil heating \cite{Jain_2016} remains a major obstacle. Proposed strategies to overcome this challenge rely on either switching off the trapping light \cite{ORI_2011_july,ORI_2011_nov,Bateman_2014,Neumeier_2024,MRL_2024,Casulleras_2024} or implementing hybrid control schemes with electric and magnetic potentials 
\cite{ORI_2012,Millen_2015,Fonseca_2016,Slezak_2018,Conangla_2020,Bykov_2022,MGL_2023,Hofer_2023,Melo_2024,Bonvin_2024_june,Bonvin_2024_nov,Hansen_2025}.

An alternative route is to employ all-electrical traps in ultra-high vacuum \cite{Bykov_2019,Goldwater_2019,Dania_2019,Bullier_2020,Martinetz_2020,Dania_2021}, where nanoparticles can achieve record-low damping rates \cite{Dania_2024}. 
The recent demonstration of co-trapping a nanoparticle together with one or more ions \cite{Bykov_2025} opens the door to leveraging the control capabilities of ion quantum optics for levitated optomechanics. For example, Doppler-cooled ions could be used to sympathetically cool the nanoparticle through their Coulomb interaction \cite{Seberson_2020,Penny_2023,Bykov_2023}. This possibility, which according to classical estimations is reachable in current setups~\cite{Bykov_2025}, would relax the requirement to optically address the nanoparticle. Furthermore,  the internal qubit transition of the ions could be directly used to prepare non-Gaussian motional states \cite{Pflanzer_2013,Toro_2021,Schut_2025}. To fully harness these capabilities, however, a quantum theory of the coupled ion-nanoparticle dynamics is needed.

In this article we provide such a quantum theory and quantify the nanoparticle motional temperatures reachable by ion-based sympathetic cooling. First, in Sec.~\ref{sec:quantum-dynamics-dual-frequency-trap}, we derive the secular motional frequencies of a nanoparticle and an ion in a dual-frequency linear Paul trap, their classical trajectories including the first motional sideband, and their quantum Hamiltonian. In Sec.~\ref{sec:coupled-quantum-dynamics}, we derive a master equation including the linearised Coulomb coupling between ion and nanoparticle as well as the common sources of dissipation in current experiments. We proceed in Sec.~\ref{sec:sympathetic-cooling} by computing analytically the steady-state occupations of the nanoparticle motional degrees of freedom under continuous Doppler cooling of the ion, as well as the corresponding ion-induced motional cooling rates. We also include the effect of  micromotion in our analysis. We extend this formalism to multiple ions in Sec.~\ref{sec:sympathetic-cooling-N}. Finally, our conclusions are presented in Sec.~\ref{sec:conclusion}.

\begin{figure}[h!]
    \centering
    \includegraphics[width=0.96\linewidth]{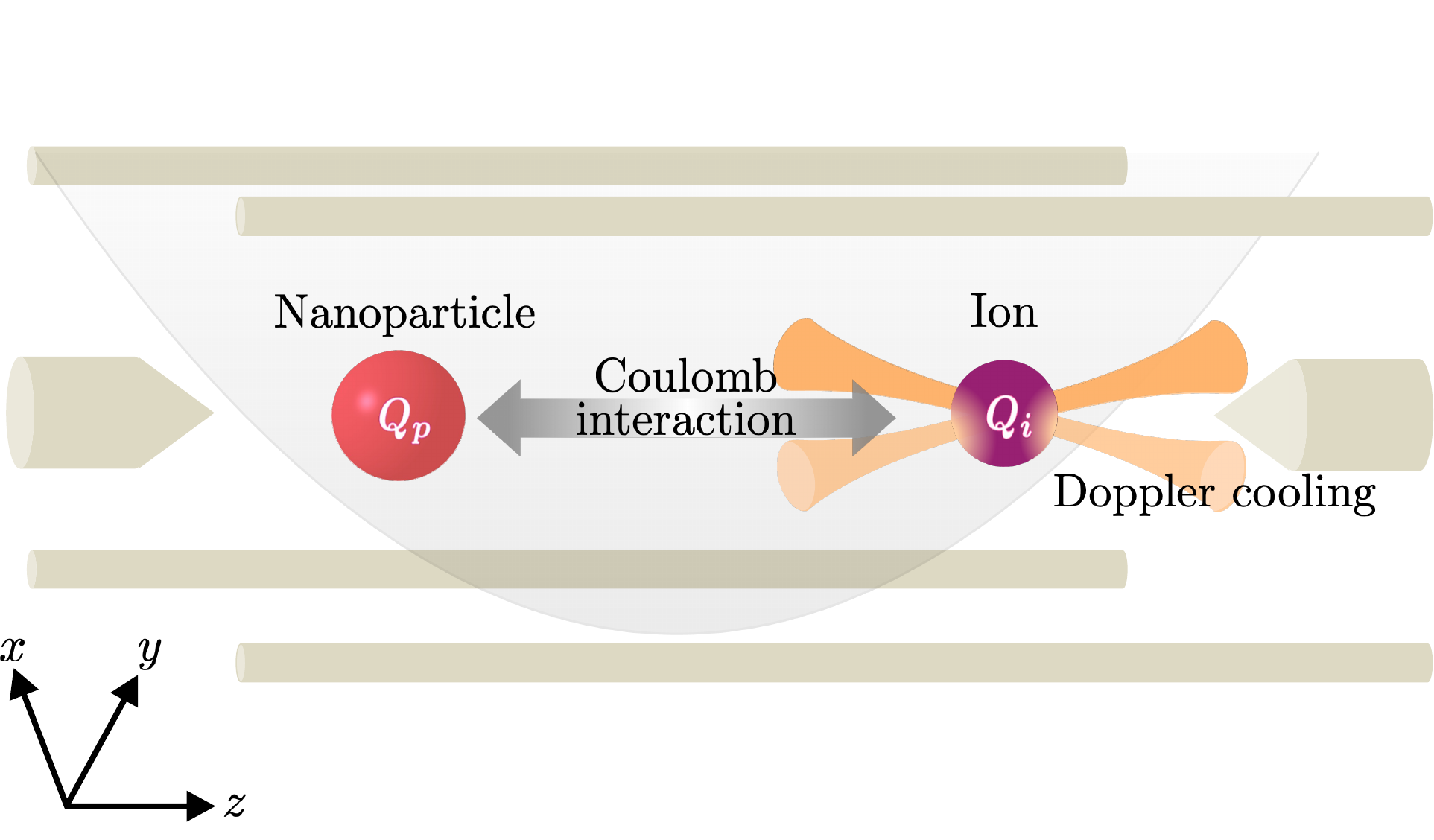}
    \caption{Scheme of the system under study. A charged dielectric nanoparticle is co-trapped with a single ion or multiple ions in a two-tone linear Paul trap. The center-of-mass motion of the nanoparticle is coupled to the ion motion by Coulomb interaction. The ion motion can be continuously Doppler-cooled using laser fields.
    }
    \label{fig:schematic}
\end{figure}

\section{Dynamics of a charged particle in a dual-frequency trap}\label{sec:quantum-dynamics-dual-frequency-trap}

The system under study, schematically depicted in Fig.~\ref{fig:schematic}, consists of a spherical dielectric nanoparticle with charge $Q_p$, radius $R_p$, and mass $M_p$, and an ion with charge $Q_i$ and mass $M_i$. 
Both the nanoparticle and the ion lie near the center of a linear Paul trap and  interact via Coulomb forces. The geometric center of the trap is chosen as the origin of the coordinate system, with the endcap electrodes lying along the $z$-axis and the four axial electrodes intersecting either the $x$- or the $y$-axis. 
Throughout this work we will consider a two-tone linear Paul trap, i.e., a trap where two AC voltages with different frequencies are applied to the electrodes. These two tones allow two objects with very different charge-to-mass ratios to be stably confined~\cite{Dehmelt_1995, Trypogeorgos_2016, Leefer_2016}.
Finally, we assume that the center of charge of the nanoparticle coincides with its center of mass, so that the motional and rotational dynamics decouple. {\color{black}This is justified in Appendix~\ref{Appendixmultipoles}, where we estimate the translation-rotation coupling and show it can be neglected for our parameters.} In this section we describe the independent classical and quantum dynamics of each particle, as a starting point for the description of their coupled quantum dynamics in subsequent sections. First, we determine the secular frequencies and classical motional trajectories of the nanoparticle and ion in a dual-frequency linear Paul trap in Sec.~\ref{sec:classical-dynamics-dual-frequency-trap}. Then, in Sec.~\ref{sec:quantum-hamiltonian-dual-frequency-trap}, we use these results to derive their quantum Hamiltonian.

\subsection{Classical dynamics in the two-tone linear Paul trap}\label{sec:classical-dynamics-dual-frequency-trap}
To derive the free quantum motional Hamiltonians of the ion and the nanoparticle, we first need to solve their classical dynamics \cite{Leibfried_2003}. We thus consider a single particle with charge $Q$ and mass $M$ in the two-tone linear Paul trap described above. The total electric potential experienced by a charged particle at position $\mathbf{R}$ near the center of the trap is given by{\color{black}~\footnote{{\color{black}We neglect higher order terms because the nanoparticle and the ion are close to the trap center. The lowest order corrections to Eq.~\eqref{potential0} are~\cite{Tabor2012, Deng2015} $\phi_{\rm corr}= C_LR_z^4/d_z^4+C_C R_z^2(R_x^2+R_y^2)/(d_zd_r)^2+C_R (R_x^2+R_y^2)^3/d_r^6$. The terms $C_{L,C,R}$ arise from Taylor expansion; $C_{L,C}\propto U_j/d_j^4$ and $C_T\propto U_j/d_j^6$, where $U_j, d_j$ are applied voltages and distances to electrodes. The relative corrections are thus $\phi_{\rm corr}/\phi\sim \frac{R_z^2}{d_z^2}, \frac{\rho^4}{d_r^4}, \frac{\rho R_z}{d_rd_z}$ due to longitudinal, transverse and cross-terms, respectively. Since $R_p/d_{z, xy}\sim10^{-4}$, corrections due to nonzero particle size are negligible, while the correction due to the ion-nanoparticle separation is $\phi_{\rm corr}/\phi\sim 10^{-3}$ for typical separations of $\sim 50\,\rm\mu m$.}}}\cite{Leibfried_2003,Berkeland_1998}
\begin{multline}\label{potential0}
    \phi \left(\bm{R},t \right) = \frac{\alpha_{\rm DC}U_{0z}}{2d_z^2}\left(2R_z^2-R_x^2-R_y^2\right)+
    \frac{\alpha_{\rm AC}}{2 d_r^2}
    \\
    \times\left(R_x^2-R_y^2\right)\left(U_{0xy}+U_s\cos(\omega_s t)+ U_f\cos(\omega_f t)\right),
\end{multline}
where $U_{0z}$ is the amplitude of the DC voltage applied to the endcaps, $U_{s}$ and $U_{f}$ are the amplitudes of the slow-oscillating and fast-oscillating voltages applied to the axial electrodes, and $U_{0xy}$ represents an offset voltage applied also to the axial electrodes. The slow and fast voltages oscillate at frequencies $\omega_s$ and  $\omega_f$ respectively. The geometry of the trap is encoded in the parameters $d_{r}$ and $d_{z}$, namely, the distances between the trap center and and the axial and endcap electrodes, respectively, and the geometric factors $\alpha_{\rm AC}$ and $\alpha_{\rm DC}$, which account for deviations from a perfect quadrupole trap (for which $\alpha_{\rm AC}=\alpha_{\rm DC}=1$).

Since the potential Eq.~\eqref{potential0} is separable, the dynamics of the particle along each axis $j=x,y,z$ are decoupled, and obey the two-tone Mathieu equation
\begin{multline}\label{eq:two-tone-mathieu-eq}
    \dv[2]{R_j(\tau)}{\tau} +\\+ \left( a_j + 2 q_{sj} l^2 \cos\left( 2 l \tau \right) + 2 q_{fj} \cos\left( 2 \tau \right) \right) R_j(\tau) = 0,
\end{multline}
where we have used for simplicity the dimensionless time variable $\tau \equiv \omega_f t/2$ and defined the ratio between slow and fast frequencies, $l \equiv \omega_s/\omega_f$. The linear Paul trap parameters are
\begin{equation}
    a_z = \frac{8 Q}{M \omega_f^2}\left(\frac{\alpha_{\rm DC}U_{0z}}{d_z^2}\right) \hspace{0.3cm};\hspace{0.3cm} q_{sz}=q_{fz}=0,
\end{equation}
\begin{equation}
    a_x =-\frac{a_z}{2}+a_{0xy} \hspace{0.4cm};\hspace{0.4cm}a_y=-\frac{a_z}{2}-a_{0xy},
\end{equation}
with $a_{0xy}\equiv 4 Q \alpha_{\rm AC} U_{0xy}/(Md_r^2\omega_f^2)$, and
\begin{equation}
    q_{fx} = -q_{fy}=\frac{a_{0xy}}{2U_{0xy}}U_f,
\end{equation}
\begin{equation}
    q_{sx} = -q_{sy}=\frac{a_{0xy}}{2U_{0xy}l^2}U_s.
\end{equation}
For certain regimes of values of the above parameters, Eq.~\eqref{eq:two-tone-mathieu-eq} admits quasi-harmonic solutions, namely, solutions that oscillate at a new frequency, known as the secular frequency, which is much smaller than $\omega_f$. These solutions also have small contributions that oscillate at the RF frequencies, referred to as micromotion. Our goal is to obtain approximate expressions for these stable solutions as well as for the secular frequencies in the experimentally relevant regime $a_j,q_{sj}^2 l^4,q_{fj}^2 \ll 1$. Since the equations of motion for each motional coordinate $R_j$ are decoupled, for the remainder of this section we will omit the sub-index $j$ for simplicity.

\subsubsection{Secular frequencies}\label{sec:secular-frequencies}

We first start by determining the secular frequencies. Note that, to trap objects with such different charge-to-mass ratios as an ion and a $\sim100\mathrm{nm}$ nanoparticle, the two RF frequencies must be very different \cite{Bykov_2025}, i.e., $\omega_s \ll \omega_f$, or equivalently, $l\ll 1$. We make use of the modified Lindstedt-Poincaré method \cite{Amore_2003, Saxena_2018} to compute the secular frequency in this regime as a perturbative expansion in the parameters $q_s l^2$ and $q_f$. We will hereafter assume that the trap voltages fulfill  $U_s < U_f$ so that one of the two perturbation parameters is at most of the same order as the second, $q_s l^2 = (U_s/U_f) q_f \lesssim q_f$. This allows us to expand in the single parameter $q_f$. 
 
The starting point of the modified Lindstedt-Poincaré method is to assume that the solution of Eq.~\eqref{eq:two-tone-mathieu-eq} oscillates at a dimensionless secular frequency $\beta$, i.e. $R(\tau)\approx \exp(i\beta \tau)$, that can be perturbatively expanded as 
\begin{align}\label{eq:beta-expansion}
    \beta^2 = a - \sum_{n=1}^\infty q_f^n \beta_{(n)}.
\end{align}
Similarly, the displacement can be expanded as
\begin{align}\label{eq:r-expansion}
    R (\tau) = \sum_{n=0}^\infty q_f^n {R}_{(n)}(\tau). 
\end{align} 
We then introduce Eqs.~(\ref{eq:beta-expansion}-\ref{eq:r-expansion}) into Eq.~\eqref{eq:two-tone-mathieu-eq} and isolate the terms of the same order in $q_f$ into independent equations. Each of them is a differential equation for ${R}_{(n)}(\tau)$ which depends on lower order terms ${R}_{(n)}(\tau),{R}_{(n-1)}(\tau),...{R}_{(0)}(\tau)$, providing a recursive method to compute the solution up to the desired order in $q_f$. Since 
the solution $R(\tau)$ is stable by assumption, any resonant driving terms $\propto \exp(\pm i\beta\tau)$ appearing in each of these equations must cancel. This condition fixes the values of the coefficients  $\beta_{(n)}$. We include terms up to second order in $q_f$ in Eq.~\eqref{eq:beta-expansion}, obtaining the following biquadratic equation for the dimensionless frequency $\beta$,

\begin{equation}\label{eq:exact-dimensionless-frequency}
    \beta^2 = a + \frac{q_f^2}{2} + \frac{q_s^2 l^4}{2 \left(l^2 - \beta^2\right)}.
\end{equation}
In deriving the above expression we have used the fact that in the regime of interest $a,q_f^2 \ll 1$, the dimensionless secular frequency must also be small, $\beta^2 \ll 1$\footnote{This is fulfilled for the parameters in Table~\ref{tab:parameters}, for which $\{a,q_f^2,\beta^2\}\sim\{10^{-8}, 10^{-12},10^{-8}\}$ for the particle and $\{a,q_f^2,\beta^2\}\sim\{10^{-5}, 0.38,0.18\}$ for the ion.}.
The true secular frequency is given by
$\Omega = \beta \omega_{f}/2$. Note that in the limit of a single-frequency linear Paul trap, $l\to 0$, Eq.~\eqref{eq:exact-dimensionless-frequency} recovers the well-known expression \cite{Leibfried_2003}
\begin{align}\label{eq:Omegafdefinition}
    \lim_{l\to 0} \Omega &= \Omega_f \equiv \frac{\omega_f}{2}\sqrt{a+ q_f^2/2}.
\end{align}

\begin{table*}[t]
\caption{\label{tab:parameters}
System parameters used as case study. These parameters correspond to the experiment in Ref.~\cite{Bykov_2025}. The horizontal line separates input / experimentally measured values from the values derived using our theoretical model.}
\begin{ruledtabular}
\begin{tabular}{lll}
\textbf{Parameter} & \textbf{Description} & \textbf{Value} \\
\hline
$R_p$ & Nanoparticle radius & $134 \times 10^{-9}~\mathrm{m}$ \\
$M_p$ & Nanoparticle mass & $2.0 \times 10^{-17}~\mathrm{kg}$ \\
$Q_p$ & Nanoparticle charge & $750 \times 1.6 \times 10^{-19}~\mathrm{C}$ \\
$M_i$ & Ion mass & $6.63 \times 10^{-26}~\mathrm{kg}$ \\
$Q_i$ & Ion charge & $1.6 \times 10^{-19}~\mathrm{C}$ \\
$[d_r,\,d_z]$ & Distances between electrode and trap center & $[0.9,\,1.7] ~\mathrm{mm}$\\
$[\alpha_{\rm AC},\,\alpha_{\rm DC}]$ & Geometric factors  & $[0.93,\,0.19]$\\
$[U_{0z},U_s,U_f,U_{0xy}]$ & Voltage amplitudes & $[56.5,80,1350,0.68] ~\mathrm{V}$\footnotemark
\\
$\omega_s$ & Slow frequency & $2\pi \times 7  ~\mathrm{kHz}$
\\
$\omega_f$ & Fast frequency & $2\pi \times 17.5  ~\mathrm{MHz}$\\
$T$ & Temperature of surrounding gas & $300~\mathrm{K}$ \\
$P$ & Pressure of surrounding gas & $7 \times 10^{-11}~\mathrm{mbar}$ \\
$\dot{E}^{\rm dop}$ & Energy heating rate of the ion due to photon recoil & $3.8 \times 10^{-22} ~\mathrm{J/s}$ \\
$\gamma^{\text{dop}}$ &  Doppler cooling rate of the ion & $2\pi \times 10~\mathrm{kHz}$
\\
$\dot{E}^{\rm td}$ & Energy heating rate of the nanoparticle due to additional noise & $2.8 \times 10^{-26}~\mathrm{J/s}$
\\
$[\Omega_{xp},\, \Omega_{zp}]$ & Mechanical frequencies of nanoparticle (measured) & $2\pi \times [1.5,\, 1] ~\mathrm{kHz}$ \\
$[\Omega_{xi},\,  \Omega_{zi}]$ & Mechanical frequencies of ion (measured) & $2\pi \times [4,\,  0.8] ~\mathrm{MHz}$ \\
$\bm{D}_i-\bm{D}_p$ &  Nanoparticle-ion equilibrium separation (measured) & $[0,\, 0,\, 51] ~\mathrm{\mu m}$ \\
\hline
$[\Omega_{xp},\, \Omega_{yp},\, \Omega_{zp}]$ & Mechanical frequencies of nanoparticle (predicted) & $2\pi \times [1.38,\, 1.27,\, 1.06] ~\mathrm{kHz}$ \\
$[\Omega_{xi},\, \Omega_{yi},\, \Omega_{zi}]$ & Mechanical frequencies of ion (predicted) & $2\pi \times [3.77,\, 3.76,\, 0.67] ~\mathrm{MHz}$ \\
$\bm{D}_i-\bm{D}_p$ &  Nanoparticle-ion equilibrium separation (predicted) & $[0,\, 0,\, 52.6] ~\mathrm{\mu m}$ \\
$\gamma^{\rm gas}$ & Damping rate of the nanoparticle due to surrounding gas & $2\pi \times 8.4 ~\mathrm{nHz}$  
\end{tabular}
\end{ruledtabular}
\footnotetext[\value{mpfootnote}]{%
\justifying
Note that the potential expressed in Eq.~\eqref{potential0} does not include an RF contribution along the $z$-axis. A small contribution was present in the experiments of Ref.~\cite{Bykov_2025} but can be minimized in future experiments by using a symmetric trap drive~\cite{Teller_2023} or a linear Paul trap with RF electrode spacing much smaller than that of the endcap electrodes. In the first approach, an RF null is achieved along the $z$-axis (that is, not only at the origin) by driving the two RF electrode pairs
 with equal amplitude and frequency, but phase-shifted by $\pi$. 
 In the second approach, an RF null is achieved only at the origin, but the RF field remains weak at all other points along the $z$-axis, minimally affecting the dynamics of the trapped objects. For example, this configuration is used in state-of-the-art trapped-ion quantum information processing experiments without compromising qubit operations mediated by collective ion motion~\cite{Pogorelov_2021}.
}
\end{table*}

The full expression for the secular frequencies obtained by solving Eq.~\eqref{eq:exact-dimensionless-frequency} is
\begin{equation}\label{eq:secular-frequency}
    \Omega_{i, p}\equiv \Omega^{\pm} = \sqrt{\frac{1}{2}\Bigg(\Omega_f^2+\frac{\omega_s^2}{4}\pm\sqrt{\left(\Omega_f^2-\frac{\omega_s^2}{4}\right)^2-\frac{q_s^2\omega_s^4}{8}}\Bigg)}.
\end{equation}
The existence of two solutions for $\Omega$ can be physically interpreted by noting that without the RF drive, e.g., for motion along the $z$ axis, these two solutions correspond to
\begin{equation}\label{eq:secular-frequency-no-modulation}
    \Omega^{\pm}=\frac{\omega_f}{2}\sqrt{a+l^2\pm |a-l^2|}=\begin{cases}
(\omega_f/2)\sqrt{a} \\
\omega_s/2.
\end{cases}
\end{equation}
The first solution is just the expected frequency given by the time-independent harmonic potential, while the second ($\omega_s/2$) corresponds to the edge of the Floquet stability zone for infinitesimally weak driving. This solution corresponds to unstable or near-unstable dynamics. Indeed, note that for small two-tone driving this solution 
becomes $\Omega \approx (\omega_f/2)(1 +
\eta q_sl^2/4)$ or equivalently $\beta=l(1+\eta q_sl^2/4)$, where $\eta = \text{sign}[l^2-a]$. Since in this case $\beta$ is very close to $l$, the third term in Eq.~\eqref{eq:exact-dimensionless-frequency} diverges, indicating a breakdown of the perturbative expansion Eq.~\eqref{eq:beta-expansion}. This solution is thus inconsistent with our initial assumption of stable oscillation and needs to be discarded. 

The physical solution of Eq.~\eqref{eq:secular-frequency} can be either $\Omega^+$ or $\Omega^-$ depending on the value of $\eta$. In our system one typically has $a\ll l^2$ for the nanoparticle and $a\gg l^2$ for the ion, so that the physical solutions correspond to  $\Omega^+$ for the ion and to $\Omega^-$ for the nanoparticle. This is a direct consequence of the hierarchy $\Omega_i\gg \omega_s\gg \Omega_p$. Note that for certain parameter choices these frequencies can be complex, corresponding to unstable behaviour~\footnote{Note that for the nanoparticle one could in principle use an approximation similar to Eq.~\eqref{eq:Omegafdefinition},  $\Omega_p\approx (\omega_f/2)\sqrt{a+q_f^2/2+(q_s^2l^2/2)}$~\cite{Leefer_2016}, which for the parameters of Table~\ref{tab:parameters} coincides with Eq.~\eqref{eq:secular-frequency} within a $~10\%$ error. However, in this approximation the unstable behaviour of Eq.~\eqref{eq:secular-frequency} is lost. Since this instability is crucial to assess the trapping dynamics we keep the full expression Eq.~\eqref{eq:secular-frequency} throughout this work.}.
The secular frequencies predicted by Eq.~\eqref{eq:secular-frequency}, shown in Table~\ref{tab:parameters} for parameters of Ref.~\cite{Bykov_2025}, quantitatively reproduce experimental observations.

\subsubsection{Displacement functions}\label{sec:displacement-functions}

The Lindstedt-Poincaré method is especially suited to computing secular frequencies, as it allows us to capture terms of all orders in the perturbative parameter $q_f$ (see, e.g.,  Eq.~\eqref{eq:Omegafdefinition}). This is not true for the computed displacement function $R(\tau)$, which remains a finite-order expansion and thus less accurate than the Floquet solutions obtained in single-frequency traps. To derive a solution with a similar level of accuracy, we employ a modified Dehmelt approximation \cite{Ghosh_1995}. First, we take the following Ansatz for the solution of Eq.~\eqref{eq:two-tone-mathieu-eq} in the regime $a,q_f^2\ll 1$,
\begin{equation}\label{AnsatzDehmelt}
    R(\tau) = e^{i\beta \tau}\big(u_0+u_s(\tau)+u_f(\tau)\big),
\end{equation}
where $u_0$ is a constant of order $1$ and $u_s(\tau),u_f(\tau)\sim O(q_f)$ are slow and fast oscillating solutions at frequencies of order $2l$ and $2$, respectively. We do not fix a specific value for $\beta$ but assume it to be small, $\beta \ll 1$. This condition is fulfilled for both the nanoparticle and the ion for the parameters given in Table~\ref{tab:parameters}, as shown in the previous section.

Introducing Eq.~\eqref{AnsatzDehmelt} into Eq.~\eqref{eq:two-tone-mathieu-eq}, we obtain
\begin{multline}\label{DehmeltAnsatzDynamicalEq}
\ddot{u}_f+\ddot{u}_s+2i\beta(\dot{u}_f+\dot{u}_s)+(a-\beta^2+2q_f\cos(2\tau)\\
+2 q_sl^2\cos(2 l \tau))(u_0+u_s+u_f)=0.
\end{multline}
We proceed by isolating the terms oscillating at the fast frequency into a single equation, where we neglect all terms of order $a q_f,q_f^2$, or $\beta^2 q_f$. The resulting approximate equation for $u_f(\tau)$ has the solution 

\begin{equation}\label{Dehmeltfast}
    u_f(\tau) \approx \frac{q_f }{2}\cos(2\tau)u_0,
\end{equation}
to lowest order in $a,q_f^2$.
 We now introduce this solution into Eq.~\eqref{DehmeltAnsatzDynamicalEq} and take a time average over one period of the fast frequency, assuming that all functions which oscillate at the slow frequency remain approximately constant within such period. We obtain
\begin{multline}
    \ddot{u}_s + 2i\beta \dot{u}_s+\frac{q_f^2 u_0}{2(1-\beta^2)}+\\+(a-\beta^2+2q_sl^2\cos(2 l\tau))(u_0+u_s)
   =0.
\end{multline}
Repeating the above steps for the slow timescale we obtain an expression for the slowly varying amplitude,
\begin{equation}\label{Dehmeltslow}
    u_s(\tau)\approx \frac{q_sl^2}{2(l^2-\beta^2)}\cos(2 l \tau)u_0.
\end{equation}
We also obtain with this method the same equation for the secular frequencies that we derived through the modified Lindstedt-Poincaré method, Eq.~\eqref{eq:exact-dimensionless-frequency}. This confirms the consistency of our secular frequency calculation. 

Combining Eqs.~\eqref{AnsatzDehmelt}, \eqref{Dehmeltfast}, and \eqref{Dehmeltslow}, 
we obtain an expression for the displacement up to a multiplicative constant $u_0$. We fix this constant by imposing the initial conditions $R(0)=1$ and $\dot{R}(0)=i\beta$, a convenient choice to write the quantum Hamiltonian, as we will see below. The final displacement in real time is
\begin{multline}\label{rclassicalDehmelt}
    R(t) = e^{i \Omega t}\left(1+\frac{q_f}{2} + \frac{q_sl^2}{2(l^2-\beta^2)}\right)^{-1}
    \\
    \times \left(1+\frac{q_f}{2}\cos(\omega_f t) + \frac{q_sl^2}{2(l^2-\beta^2)}\cos(\omega_s t)\right),
\end{multline}
where $\Omega$ is the secular frequency given in Eq.~(\ref{eq:secular-frequency}). Note that in the limit $l\to 0$ the above equation recovers the well-known expression for single-frequency traps \cite{Leibfried_2003}. Note also that the solution Eq.~\eqref{rclassicalDehmelt} is consistent with the initial assumptions -- specifically, the assumption of $u_s(\tau)$ being small --  only if
\begin{equation}\label{condpert}
    \frac{q_sl^2}{2(l^2-\beta^2)} \ll 1.
\end{equation}
This is yet a second manifestation of the fact that for $\beta\approx l$ the perturbative expansion breaks down, as discussed above.
The condition Eq.~\eqref{condpert} is fulfilled for the parameters of Table~\ref{tab:parameters} and will be assumed to hold hereafter. 

\subsection{Quantum Hamiltonian of a particle in the two-tone linear Paul trap}\label{sec:quantum-hamiltonian-dual-frequency-trap}

The Hamiltonian describing the dynamics of a particle along one of the Cartesian axes of the linear Paul trap is given in the Schr\"odinger picture by \cite{Leibfried_2003}
\begin{equation}\label{eq:time-dependent-hamiltonian}
    \hat{H}(t) =  \frac{\hat{P}^2}{2M} + \frac{M}{2} W(t) \hat{R}^2,
\end{equation}
with $\hat{R}$ and $\hat{P}$ the position and momentum operators along the chosen axis, and 
\begin{equation}\label{Wpotentialdef}
    W(t) \equiv \frac{\omega_f^2}{4} \left(a + 2 q_s l^2 \cos\left( \omega_s t \right) + 2 q_f \cos\left( \omega_f t \right) \right).
\end{equation}
Note that the function $W(t)$ is different for each of the three axes. In the following,  Heisenberg picture operators are denoted by the superindex $(H)$.

Since the Heisenberg picture position operator $\hat{R}^{(H)}(t)$ satisfies the two-tone Mathieu equation Eq.~\eqref{eq:two-tone-mathieu-eq}, we can combine it with the scalar solution obtained above, $R(t)$,
into the operator
\cite{Leibfried_2003}
\begin{align}\label{CHdef}
           \hat{C}^{(H)}(t) &\equiv \sqrt{\frac{M}{2 \hbar \Omega}} i \left( R(t) \dot{\hat{R}}^{(H)}(t) - \dot{R}(t) \hat{R}^{(H)}(t) \right),
\end{align}
which is by definition time-independent, $\hat{C}^{(H)}(t)=\hat{C}^{(H)}(0)$. Introducing the initial conditions $R(0)=1$ and $\dot{R}(0)=i\Omega$ we can write this operator as a Schr\"odinger picture ladder operator for a harmonic oscillator with the secular frequency $\Omega$, i.e.,
\begin{equation}\label{CHdefLadder}
    \hat{C}^{(H)}(t) = \sqrt{\frac{M}{2 \hbar \Omega}} i \left(  \frac{{\hat{P}}^{(H)}(0)}{M} - i\Omega \hat{R}^{(H)}(0) \right) \equiv \hat{b}.
\end{equation}
Combining Eqs.~(\ref{CHdef}-\ref{CHdefLadder}) and their complex conjugates, we can reexpress the position and momentum operators as 
\begin{equation}
    \hat{R}^{(H)}(t) = \sqrt{\frac{\hbar}{2 M \Omega}} \left( \hat{b} R^*(t) + \hat{b}^\dagger R(t) \right),
\end{equation}
\begin{equation}
    \hat{P}^{(H)}(t) = \sqrt{\frac{\hbar M \Omega}{2}} \left( \hat{b} \frac{\dot{R}^*(t)}{\Omega} + \hat{b}^\dagger \frac{\dot{R}(t)}{\Omega} \right).
\end{equation}
Substituting these expressions into the Heisenberg picture Hamiltonian, we obtain
\begin{align}\label{eq:non-rwa-hamiltonian}
    \hat{H}^{(H)}(t) &= \frac{\hbar}{4\Omega} \left( \hat{b} \bm{v}^*(t) + \text{H.c.} \right)^2,
\end{align}
where we have defined the vector
\begin{equation}
    \bm{v} (t) \equiv \mqty[ \dot{R} (t),  R(t) \sqrt{W (t)} ]^T.
\end{equation}

In the regime of interest $a,q_f \ll 1$, the above Hamiltonian can be simplified by introducing the solution 
Eq.~\eqref{rclassicalDehmelt} and undertaking a series of rotating wave approximations to eliminate all time-dependent terms. The Hamiltonian thus simplifies to that of a harmonic oscillator at the secular frequency,
\begin{equation}\label{eq:rwa-hamiltonian}
    \hat{H}^{(H)}(t)\approx\hat{H} =\hbar \Omega\hat b^\dagger\hat b.
\end{equation}
The validity conditions of the rotating wave approximations are involved in general, but can be simplified in the case $\beta^2 \ll l^2$, which is
typically fulfilled by the nanoparticle, and in the case $l^2 \ll \beta^2$, which is typically fulfilled by the ion. Specifically, the approximate Hamiltonian Eq.~\eqref{eq:rwa-hamiltonian} is valid for the nanoparticle provided that $q_f \omega_f, q_s\omega_s  \ll 16\Omega$, and for the ion provided that $q_f \omega_f \ll 16\Omega$ and $(q_s/64)(\omega_s/\Omega)^3\vert 1-(q_s/8)^2(\omega_s/\Omega)^2\vert \ll 1$. All these conditions are fulfilled for the parameters in Table~\ref{tab:parameters}.

\section{Coupled quantum dynamics of ion and nanoparticle}\label{sec:coupled-quantum-dynamics}

The coupled quantum dynamics of the motional degrees of freedom of ion and nanoparticle can be described by the quantum master equation
\begin{align}\label{eq:basic-master-eq}
    \dot{\hat{\rho}} &= -\frac{i}{\hbar} \comm{\hat{H}_{i} + \hat{H}_{p} + \hat{V}_{ip}}{\hat{\rho}} + \mathcal{D}_p \left( \hat{\rho} \right) + \mathcal{D}_i \left( \hat{\rho} \right).
\end{align}
Here, $\hat{\rho}$ is the density matrix of the nanoparticle and ion motional degrees of freedom.
The free motional Hamiltonians of the nanoparticle and the ion, $\hat{H}_i$ and $\hat{H}_p$, can be written using Eq.~\eqref{eq:rwa-hamiltonian}  as
\begin{equation}\label{Hsigmasecular}
    \hat{H}_\sigma = \frac{\hat{\mathbf{P}}^2_\sigma}{2M_\sigma}+\frac{M_\sigma}{2}\sum_{j=x,y,z}\Omega_{ j\sigma}^2\hat{R}^2_{ j\sigma}, 
\end{equation}
for for $\sigma=i,p$, 
with $\Omega_{j \sigma}$ the secular frequency along axis $j$ and with
$\hat{\bm{R}}_\sigma=[\hat{R}_{x\sigma},\hat{R}_{y\sigma},\hat{R}_{z\sigma}]$ and $\hat{\bm{P}}_\sigma=[\hat{P}_{x\sigma},\hat P_{y\sigma},\hat P_{z\sigma}]$ the respective position and momentum operators, fulfilling $[\hat{R}_{j\sigma},\hat{P}_{j'\sigma'}]=i\hbar \delta_{jj'}\delta_{\sigma\sigma'}$. 
The term $\hat{V}_{ip}$ in Eq.~\eqref{eq:basic-master-eq}  represents the Coulomb coupling between ion and nanoparticle, whereas the second and third terms describe the dissipative dynamics. In this section we discuss in detail the form of each of these terms. First, in 
Sec.~\ref{sec:coulomb-interaction}, we linearize the Coulomb interaction, determine the equilibrium ion and nanoparticle positions, and compute their motional coupling rates. Then, in Sec.~\ref{sec:dissipative-dynamics}, we describe the terms contributing to the dissipative dynamics of ion and nanoparticle.

\subsection{Coulomb interaction and dynamical stability}\label{sec:coulomb-interaction}

Both nanoparticle and ion are charged and thus interact via electrostatic forces. Assuming the ion-nanoparticle separation is much larger than the diameter of the nanoparticle, their classical interaction energy is given by the Coulomb expression for point charges,
\begin{align}\label{Coulombclassical}
    V_{ip} &= \frac{1}{4 \pi \varepsilon_0} \frac{Q_i Q_p}{\abs{\bm{R}_i - \bm{R}_p }},
\end{align}
with $\varepsilon_0$ the vacuum permittivity. We assume $Q_iQ_p>0$ so that the forces are repulsive. The combination of Coulomb repulsion and the restoring force exerted by the electrodes determines the equilibrium positions of ion and nanoparticle within the trap, which will in general not be at the origin of coordinates. The equilibrium positions, which we denote by $\bm{D}_\sigma \equiv [d_{x\sigma},d_{y\sigma},d_{z\sigma}]^T$, are determined as the points where all forces vanish,
\begin{equation}\label{eq:equilibrium-config-condition}
    \left(M_\sigma\Omega_{j\sigma}^2 R_{j\sigma}+\frac{\partial}{\partial R_{ j\sigma}}V_{ip}\right)_{\mathbf{R}_i=\mathbf{D}_i,\mathbf{R}_p=\mathbf{D}_p}=0.
\end{equation}
This condition forms a system of $6$ nonlinear equations which can have multiple solutions. For two trapped objects -- e.g., one nanoparticle and one ion -- all these solutions correspond to both objects lying along the same coordinate axes. The solution where both objects lie along the $z$-axis is given by $d_{x\sigma} = d_{y\sigma }=0$ and $d_{zp}=-d_{zi}M_i\Omega_{zi}^2/(M_p\Omega_{zp}^2)=D(1+M_p\Omega_{zp}^2/(M_i\Omega_{zi}^2))^{-1}$ with $D$ being the total distance between ion and nanoparticle,
\begin{equation}\label{separation}
    D \equiv \abs{\bm{D}_i - \bm{D}_p} = \sqrt[3]{\frac{Q_i Q_p}{4\pi\varepsilon_0}\left(\frac{1}{M_p\Omega_{zp}^2}+\frac{1}{M_i\Omega_{zi}^2}\right)}.
\end{equation}
Solutions along the $x$- and $y$-axes have the same form as the above under exchange of indices $z\leftrightarrow x$ or $z\leftrightarrow y$. We remark that some of these solutions might correspond to local maxima of the potential in coordinate space and thus to unstable equilibria. Stability of the solutions is discussed below.

We quantise the Coulomb energy in Eq.~\eqref{Coulombclassical} by promoting the position variables to operators, $\mathbf{R}_\sigma\to\hat{\mathbf{R}}_\sigma$. Since we are interested in dynamics near equilibrium positions, we linearise the Coulomb interaction around them. We first write the displacement operators as
\begin{equation}\label{dispfromeq}
    \hat R_{j\sigma} = d_{j\sigma}+\delta \hat R_{j\sigma}.
\end{equation}
The new operators $\delta \hat R_{j\sigma}$  have the same canonical momenta $ \hat P_{j\sigma}$ and describe displacements from the equilibrium position. They are assumed small, i.e., $\langle \delta \hat R_{j\sigma}\delta \hat R_{j'\sigma'}\rangle (t) \ll D^2$, so that the Coulomb interaction can be well approximated by its second-order Taylor expansion in the six variables $ \delta \hat R_{j\sigma}/D$. Under this approximation the total ion-nanoparticle Hamiltonian becomes quadratic and can be compactly written as
\begin{multline}\label{Hipquantum}
    \hat{H}_{\rm tot}\equiv \hat{H}_i + \hat{H}_p + \hat{V}_{ip} \approx 
    \frac{1}{2}\hat{\bm{P}}^T \bar{M}  \hat{\bm{P}} + \frac{1}{2} \hat{\bm{X}}^T  \bar{V} \hat{\bm{X}}.
\end{multline}
Here, we have discarded constant terms and defined the compound variable vectors
\begin{equation}\label{Xdef}
    \hat{\bm{X}} \equiv [\delta \hat R_{xi},\delta \hat R_{yi},\delta \hat R_{zi},\delta \hat R_{xp},\delta \hat R_{yp},\delta \hat R_{zp}]^T,
\end{equation}
and
\begin{equation}\label{Pdef}
    \hat{\bm{P}}\equiv [ \hat P_{xi},\hat P_{yi},\hat P_{zi},\hat P_{xp},\hat P_{yp},\hat P_{zp}]^T,
\end{equation}
the diagonal matrix of inverse masses,
\begin{align}\label{Massmat}
    \bar{M} &\equiv \mqty[
        M_i^{-1}\mathbb{1}_{3\times 3} && \mathbb{0}_{3\times 3} \\
        \mathbb{0}_{3\times 3} && M_p^{-1}\mathbb{1}_{3\times 3}
    ],
\end{align}
and the generalised potential matrix
\begin{equation}\label{Vatrixdef}
    \bar{V} \equiv 
    \left[\begin{array}{cc}
        -\bar{N}_{\rm coul} & \bar{N}_{\rm coul} \\
         \bar{N}_{\rm coul}& -\bar{N}_{\rm coul}
    \end{array}\right]+
   \left[
    \begin{array}{cc}
        \bar{V}_{di} &
        \mathbb{0}_{3\times 3} \\
        \mathbb{0}_{3\times 3} & 
        \bar{V}_{dp}
    \end{array}\right],
\end{equation}
with $\bar{V}_{di}\equiv M_i\text{diag}[\Omega_{xi}^2,\Omega_{yi}^2,\Omega_{zi}^2]$, $\bar{V}_{dp}\equiv M_p\text{diag}[\Omega_{xp}^2,\Omega_{yp}^2,\Omega_{zp}^2]$, and
\begin{equation}\label{Ncoulomb}
    \bar{N}_{\rm coul}\equiv  \frac{Q_i Q_p}{4 \pi \varepsilon_0D^3} \!\left[\mathbb{1}_{3\times 3}-3\frac{(\bm{D}_i-\bm{D}_p)\!\otimes\!(\bm{D}_i-\bm{D}_p)}{D^2}\right]
\end{equation}
with $x\otimes y$ denoting the Kronecker product.

The potential matrix Eq.~\eqref{Vatrixdef} is in general not diagonal, resulting in coupling between the motional degrees of freedom. The diagonal terms of $\bar{V}$ describe the new motional frequencies, which are renormalised due to Coulomb interaction. These new frequencies read
\begin{equation}\label{frequencyrenorm}
    \Omega_{j\sigma}^{'2} \equiv \Omega_{j\sigma}^2 + \frac{Q_i Q_p}{4 \pi \varepsilon_0 M_\sigma D^3}\left(3\frac{(d_{ji}-d_{jp})^2}{D^2}-1\right).
\end{equation}
Both the frequency renormalisations and the motional couplings induced by Coulomb interaction can result in the Hamiltonian for the chosen equilibrium positions  $\bm{D}_i$ and $\bm{D}_p$ not being dynamically stable, even if the uncoupled dynamics of ion and nanoparticle are. A quadratic Hamiltonian -- both classical and quantum -- is dynamically stable if the matrix
\begin{equation}\label{Kdef}
    \bar{K}\equiv  \mqty[ \mathbb{0}_{6\times 6} & -\mathbb{1}_{6\times 6} \\ \mathbb{1}_{6\times 6} & \mathbb{0}_{6\times 6}] \mqty[\bar{M} &  \mathbb{0}_{6\times 6}\\ \mathbb{0}_{6\times 6}& \bar{V}]
\end{equation}
is diagonalisable and has purely imaginary eigenvalues \cite{Lanchares_2014, Kustura_2019}. Hereafter we focus on equilibria that fulfill the above stability criterion. We remark that this criterion only establishes stability of the secular solutions of the nanoparticle and ion dynamics, but the full solution of the dynamics including micromotion can still be unstable. For the results of this work full stability has also been confirmed through a more stringent stability criterion based on Floquet theory, which is introduced in Sec.~\ref{sec:micromotion}.

As discussed above, all the equilibrium configurations correspond to ion and nanoparticle lying along the same Cartesian axis. In this case the matrix $\bar{N}_{\rm coul}$ is diagonal, simplifying the dynamics in two ways. First, the motional degrees of freedom of the ion along each Cartesian axis are uncoupled from each other, and the same is true for the nanoparticle. Second, the only interactions between ion and nanoparticle are of the form $\sim \hat{R}_{ji}\hat{R}_{jp}$ for $j=x,y,z$, i.e., only the motional degrees of freedom along the same axis couple to each other. As a result the system can be decomposed into three decoupled subsystems, each formed by two harmonic oscillators. We use this result in the following sections to obtain analytical expressions for the steady-state expectation values.

To compute the motional coupling rates between ion and nanoparticle we define adimensional position and momentum quadratures,
\begin{equation}
    \delta \hat{R}_{j\sigma}\equiv R_{j\sigma}^{\rm zpf}\hat{q}_{j\sigma}\equiv \sqrt{\frac{\hbar}{2 M_\sigma \Omega_{j\sigma}'}}\hat{q}_{j\sigma},
\end{equation}
\begin{equation}
    \hat{P}_{j\sigma}\equiv P_{j\sigma}^{\rm zpf}\hat{p}_{j\sigma}\equiv \sqrt{\frac{\hbar M_\sigma \Omega_{j\sigma}'}{2 }}\hat{p}_{j\sigma},
\end{equation}
where $R_{j\sigma}^{\rm zpf}$ and $P_{j\sigma}^{\rm zpf}$ are the zero-point displacement and momentum in the renormalised harmonic trap, respectively. The newly defined operators commute as $[\hat{q}_{j\sigma},\hat{p}_{j'\sigma'}]=2i\delta_{jj'}\delta_{\sigma\sigma'}$. In terms of these operators the Hamiltonian Eq.~\eqref{Hipquantum} is
\begin{equation}\label{eq:quantum-hamiltonian-1d}
    \hat{H}_{\rm tot} = \sum_{\substack{j=x,y,z \\ \sigma=i,p}} \frac{\hbar \Omega_{j\sigma}'}{4} \left( \hat{q}_{j\sigma}^2 + \hat{p}_{j\sigma}^2 \right) + \sum_{j=x,y,z} \hbar g_{j} \hat{q}_{ji}\hat{q}_{jp},
\end{equation}
where the coupling rates are given by
\begin{equation}
    g_{z} = -\frac{1}{2 \pi \varepsilon_0}\frac{Q_i Q_p}{ D^3} \frac{R_{zi}^{\rm{zpf}} R_{zp}^{\rm{zpf}}}{\hbar},\label{gz}
\end{equation}
\begin{equation}
    g_{j} = \frac{1}{4 \pi \varepsilon_0}\frac{Q_i Q_p}{D^3} \frac{R_{ji}^{\rm{zpf}} R_{jp}^{\rm{zpf}}}{\hbar }, \quad (j = x,y).\label{gr}
\end{equation}

Fig.~\ref{fig:g,D-vs-Qp,Mp} shows the total equilibrium separation between ion and nanoparticle, $D$, as well as the radial and longitudinal coupling rates, Eqs.~(\ref{gz}-\ref{gr}), as a function of nanoparticle charge and mass. We use the experimentally relevant parameters of Table~\ref{tab:parameters} and focus on the equilibrium configuration where both nanoparticle and ion lie along the $z$-axis, which for these parameters is the only dynamically stable equilibrium. 
As shown in Fig.~\ref{fig:g,D-vs-Qp,Mp}(a), $D$ increases with nanoparticle charge as the stronger Coulomb repulsion pushes ion apart from nanoparticle. As a consequence, the coupling rates are decreasing functions of charge $Q_p$. 
The coupling rates $g_x$ and $g_y$ diverge at $Q_p\approx 170~\mathrm{e}$ and $Q_p\approx 260~\mathrm{e}$, where the renormalised motional frequencies $\Omega_{xp}'$ and $\Omega_{yp}'$ become imaginary, indicating an instability. Mathematically, the divergence arises from the zero-point motion becoming unbounded as the frequency approaches zero, but in a physical setting the micromotion would induce Floquet instability before this divergence occurs. A second instability occurs for $Q_p\sim 10^3~\mathrm{e}$ where the bare frequencies Eq.~\eqref{eq:secular-frequency} become complex. This instability is simply a consequence of using fixed trap parameters while increasing charge to mass ratio. 

The separation $D$ is practically constant with mass $M_p$ within the range shown in Fig.~\ref{fig:g,D-vs-Qp,Mp}(b), as it is dominated by the factor $(M_i\Omega_{zi})^{-2}\gg (M_p\Omega_{zp})^{-2}$ (see Eq.~\eqref{separation}). Within the mass range shown in the figure, the $g_z$ coupling rate decreases with $M_p$ due to the decrease in the zero-point fluctuations along this axis, 
$R_{z p}^{\rm zpf}\sim (M_p)^{-1/4}$. 
In contrast, the coupling rates along the $x$- and $y$-axes increase with mass as the renormalised frequencies along these axes become smaller. In analogy to the dependency with $Q_p$, the frequencies $\Omega_{xp}'$ and $\Omega_{yp}'$ eventually become imaginary due to this renormalisation, at $M_p\approx 8.6\times10^{-17}~\mathrm{kg}$ and $M_p\approx 5.7\times10^{-17}~\mathrm{kg}$, respectively. The other stability edge at $M_p\approx1.5\times 10^{-17}~\mathrm{kg}$ again corresponds to the frequencies Eq.~\eqref{eq:secular-frequency} becoming imaginary, as the charge-to-mass ratio becomes too big for the given trap parameters.

\begin{figure}[t!]
    \centering
    \includegraphics[width=1\linewidth]{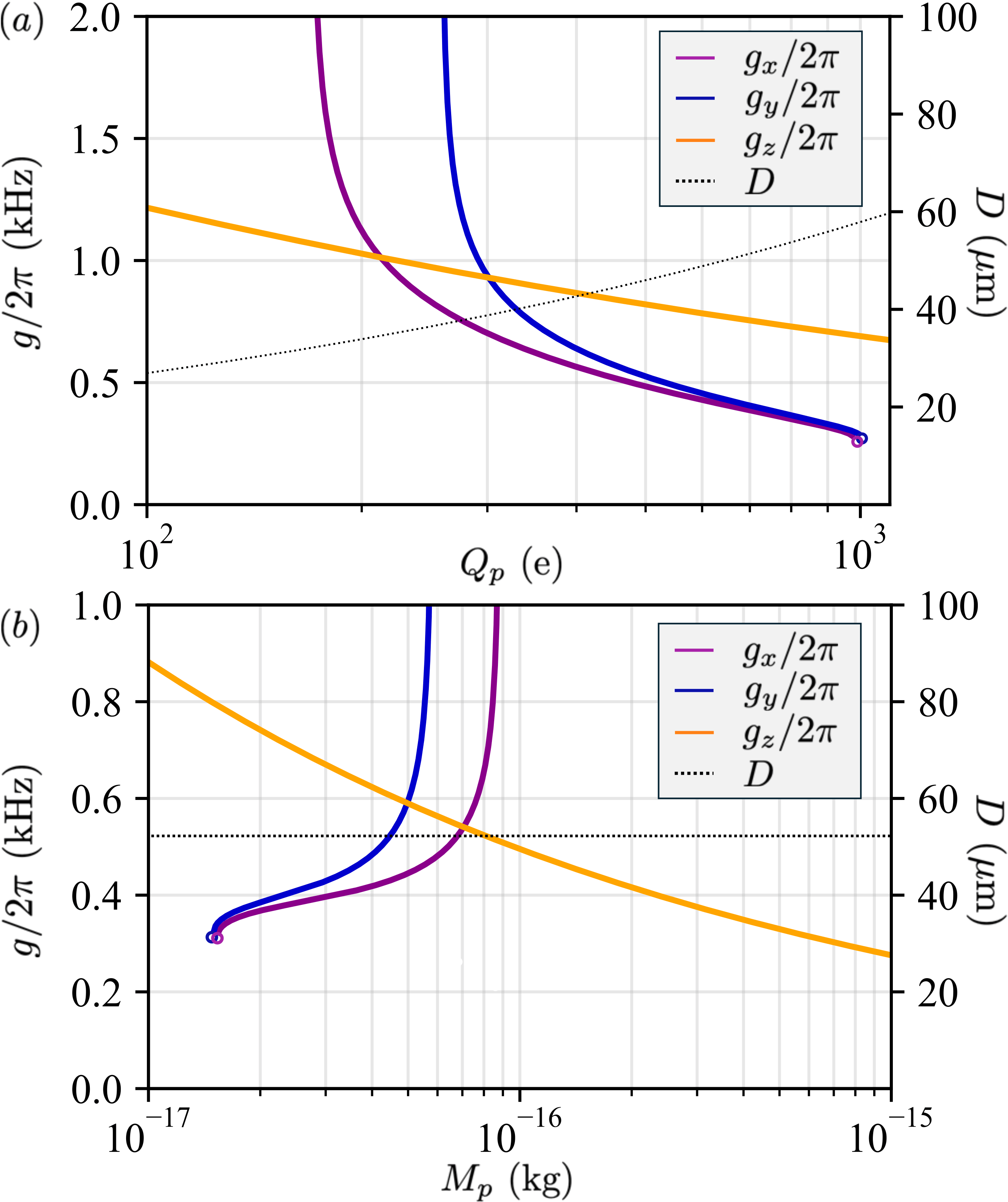}
    \caption{Coupling rate between ion and nanoparticle center-of-mass motion along the $x$- (Eq.~\eqref{gr}, magenta), $y$- (Eq.~\eqref{gr}, blue), and $z$-axis (Eq.~\eqref{gz}, orange) and equilibrium distance between the ion and nanoparticle (Eq.~\eqref{separation}, dashed), versus nanoparticle charge $Q_p$ (panel a) and nanoparticle mass $M_p$ (panel b), for the parameters of Table~\ref{tab:parameters}. The equilibrium position corresponds to nanoparticle and ion lying along the $z$-axis. The open dots mark the points beyond which the coupled system becomes unstable; see main text for details.}
    \label{fig:g,D-vs-Qp,Mp}
\end{figure}

\subsection{Dissipative dynamics}\label{sec:dissipative-dynamics}

Let us finally describe the dissipative dynamics experienced by nanoparticle and ion, namely, the general dissipators $\mathcal{D}_p \left( \hat{\rho} \right)$ and $\mathcal{D}_i \left( \hat{\rho} \right)$ in Eq.~\eqref{eq:basic-master-eq}, respectively. {\color{black}Since $g_j/|\Omega_{ji}'-\Omega_{jp}'|\ll 1$ in our system, the motional modes of the ion and the nanoparticle hybridize very weakly, and we can safely use the local master equation approach, i.e., just add the ion and nanoparticle dissipators independently~\cite{Cattaneo_2019}.} The nanoparticle experiences three sources of dissipation,
\begin{equation}\label{dissipatorP}
    \mathcal{D}_p \left( \hat{\rho} \right)=\sum_{j=x,y,z}\left(\mathcal{D}_{j,\rm gas} \left( \hat{\rho} \right)+\mathcal{D}_{j, \rm fb} \left( \hat{\rho} \right)
    +
    \mathcal{D}_{j,\rm td} \left( \hat{\rho} \right)\right).
\end{equation}
The first contribution is decoherence due to collisions with surrounding gas, and can be approximated as
\begin{multline}\label{gasdamping}
    \mathcal{D}_{j,\rm{gas}} \left( \hat{\rho} \right) \approx  \left( \Gamma^{\rm{gas}}_j + \gamma^{\rm gas} \right) \mathcal{L}_{\hat{b}_{jp},\hat{b}_{jp}^\dagger} \left( \hat{\rho} \right) +\\+ \Gamma^{\rm{gas}}_j\mathcal{L}_{\hat{b}_{jp}^\dagger,\hat{b}_{jp}} \left( \hat{\rho} \right),
\end{multline}
where we define the Lindblad dissipator $\mathcal{L}_{\hat{A},\hat{B}} \left( \hat{\rho} \right) \equiv \hat{A} \hat{\rho} \hat{B} - \frac{1}{2} \{\hat{B} \hat{A},\hat{\rho}\}$~\cite{Diosi_1995,CGB_2019}, with the curly brackets denoting the anti-commutator, and where $\hat b_{j\sigma}=(\hat q _{j\sigma} +i\hat p_{j\sigma})/2$ are the motional ladder operators of particle $\sigma=i,p$ along the coordinate axis $j=x,y,z$, defined in Eq.~\eqref{CHdefLadder}. The heating rate due to the surrounding gas is $\Gamma_j^{\rm gas}\equiv \gamma^{\rm gas}k_B T/(\hbar \Omega'_{jp})$,
with $k_B$ being Boltzmann's constant and $T$ the temperature, and the damping rate is~\cite{Li_2011, Beresnev_1990}
\begin{align}
    \gamma^{\rm gas} = 0.619 \frac{6 \pi R_p^2}{M_p} P \sqrt{\frac{2 m_o}{\pi k_B T}},
\end{align}
where $P$ is the pressure of the gas and $m_o$ is the molecular mass of the gas (which we fix to the mass of hydrogen molecules, $m_o = 3.35 \times 10^{-27}~\mathrm{kg}$). Equation~\eqref{gasdamping} is valid provided that $k_B T \gg \hbar \Omega_{jp}'$ and $\Gamma_j^{\rm gas} \ll 2\Omega_{jp}'$, which are fulfilled at room temperature and the ultra-high vacuum of typical experiments \cite{Dania_2024, Bykov_2025}.

The second term in Eq.~\eqref{dissipatorP} describes measurement-based feedback cooling of the nanoparticle, which is commonly employed to stabilise the nanoparticle in electrical levitation experiments~\cite{Millen_2015, Slezak_2018,Bykov_2022,MGL_2023,Hofer_2023,Melo_2024,Bonvin_2024_june,Bonvin_2024_nov,Hansen_2025,Goldwater_2019,Dania_2019,Bullier_2020,Dania_2021,Dania_2024}. We assume the measurement is performed optically and thus model the effect of feedback through the dissipator{\color{black}~\cite{Wiseman_1993}}
\begin{align}\label{feedbackDissipator}
    \mathcal{D}_{j,\rm{fb}} \left(\hat{\rho}\right) &= \gamma^{\rm{fb}} \mathcal{L}_{\hat{b}_{jp}, \hat{b}_{jp}^\dagger} \left(\hat{\rho}\right) - \frac{\Gamma^{\rm ba}_j}{2} \comm{\hat{q}_{jp}}{\comm{\hat{q}_{jp}}{\hat{\rho}}}.
\end{align}
The first term describes motional damping at a rate $\gamma^{\rm{fb}}$, which
for simplicity we assume equal for all motional axes. 
The second term describes decoherence due to measurement backaction, i.e., recoil heating. Assuming the measurement is performed with a conventional Gaussian beam, the recoil heating rate is $\Gamma^{\rm ba}_j= \zeta (P_t/W_t^2)\alpha_p^2 k_0^5(R_{jp}^{\rm{zpf}})^2/(15\pi^2\hbar\varepsilon_0^2 c)$~\cite{CGB_2019,MaurerPRA2023}, where $P_t$ and $W_t$ are the power and waist of the probing beam; $\alpha_p=4\pi\varepsilon_0R_p^3(\varepsilon-1)/(\varepsilon+2)$ is the polarizability of the nanoparticle, with $\varepsilon$ its relative permittivity; $k_0$ is the wavenumber of the probing beam; and $c$ is the speed of light. The factor $\zeta\in\{1,2,7\}$ depends on the direction and polarisation of the optical probe beam with respect to the motional axis $j$. For ideal feedback, the damping rate is proportional to the light intensity at the nanoparticle position, i.e., we can write $\gamma^{\rm fb} = c_{\rm fb} P_t/W_t^2$, with $c_{\rm fb}$ a constant that depends on the specific experimental setup. This allows us to write the recoil heating rate as
\begin{equation}\label{Gammarecoil}
    \Gamma^{\rm ba}_j= \zeta (\gamma^{\rm fb}/c_{\rm fb})\frac{\alpha_p^2 k_0^5}{15\pi^2\hbar\varepsilon_0^2 c}\left(R_{jp}^{\rm{ zpf } }\right)^2,
\end{equation}
an expression that is convenient when particularizing to specific setups.

The third term in Eq.~\eqref{dissipatorP} describes other noise experienced by the trapped particles, e.g., stray electromagnetic fields, noise in the voltage applied to the linear Paul trap electrodes, mechanical vibrations, etc. Effectively, these noise sources all act as a stochastic displacement of the trap center, and can be described as
\begin{align}
    \mathcal{D}_{j, \rm{td}} \left(\hat{\rho}\right) 
    &= -\frac{\Gamma^{\rm{td}}_j}{2} \comm{\hat{q}_{jp}}{\comm{\hat{q}_{jp}}{\hat{\rho}}},
\end{align}
with noise rate $\Gamma^{\rm td}_j\equiv \dot{E}^{\rm td}/(\hbar\Omega_{jp}')$. The energy heating rate $\dot{E}^{\rm td}$ is a free parameter of our model that can be fixed to experimental measurements. In current electrical levitation experiments in high vacuum this term is dominant, as $\Gamma^{\rm td}_j \gg\Gamma_j^{\rm gas}, \Gamma_j^{\rm ba}$~\cite{Dania_2024}.

Finally, we shift our focus to the motional dissipative dynamics of the ion. We assume the ion is continuously Doppler-cooled using laser fields, such that the motional dissipation is dominated by the interaction with those fields, i.e.,
\begin{multline}\label{dissipatorI}
    \mathcal{D}_i \left( \hat{\rho} \right)=\sum_{j=x,y,z}\left( \Gamma_j^{\rm{dop}} + \gamma^{\rm{dop}} \right) \mathcal{L}_{\hat{b}_{ji},\hat{b}_{ji}^\dagger} \left(\hat{\rho}\right) +\\+ \Gamma_j^{\rm{dop}} \mathcal{L}_{\hat{b}_{ji}^\dagger,\hat{b}_{ji}} \left(\hat{\rho}\right).
\end{multline}
Here, $\gamma^{\rm{dop}}$ is the Doppler cooling rate, and $\Gamma^{\rm{dop}}_j\equiv \dot{E}^{\rm dop}/(\hbar \Omega_{ji}')$ is the heating rate due to recoil heating and residual excitation of blue sideband transitions \cite{Eschner_2003}, with $\dot{E}^{\rm dop}$ being the energy heating rate.

This section concludes the derivation of the coupled motional dynamics of ion and nanoparticle. In the following sections we will use our derived master equation to explore ion-based sympathetic cooling of the nanoparticle center-of-mass motion.

\section{Sympathetic cooling}\label{sec:sympathetic-cooling}

Electrically trapped nanoparticles have lower motional frequencies than their optically levitated counterparts, which makes reaching the ground state harder. Trapped ions could help bridge this gap, as their motion can be Doppler-cooled at a fast rate. Since ion motion and nanoparticle motion are coupled via Coulomb interaction, continuously cooled ions can be used as energy sinks to remove thermal energy from the nanoparticle motion. In this section, we quantify this so-called sympathetic cooling. First, in Sec.~\ref{subsec:eomN}, we derive equations of motion for the system observables and compute the occupation of the nanoparticle center-of-mass motion in the presence of the continuously cooled ion. Then, in Sec.~\ref{sec:micromotion}, we quantify the impact of micromotion on sympathetic cooling.

\subsection{Equations of motion and phonon occupations}\label{subsec:eomN}

Since the ion motion along axis $j$ only couples to the nanoparticle motion along the same axis $j$, as discussed above, the dynamics can be described by three independent subsystems of two coupled oscillators each. The dynamics of the expectation value of an arbitrary operator in one of these subsystems, $\hat{O}_j$, can be computed directly from the master equation Eq.~\eqref{eq:basic-master-eq},
\begin{multline}\label{eomOgeneral}
    \dv{\langle\hat O_j\rangle}{t}= -\frac{i}{\hbar} \Tr[\comm{\hat{O}_j}{\hat{H}_{\rm tot}} \hat{\rho} ] +\\+ \Tr[ \hat{O}_j \left(\mathcal{D}_p \left(\hat{\rho}\right)+\mathcal{D}_i \left(\hat{\rho}\right)\right) ],
\end{multline}
where $\hat{H}_{\rm tot}=\hat{H}_{i} + \hat{H}_{p} + \hat{V}_{ip}$. Specifically, we derive the closed systems of equations
\begin{equation}
    \dv{ \langle \hat{\bm{B}}_j\rangle}{t} = \bar{A}_{j}\langle \hat{\bm{B}}_j\rangle
\end{equation}
and
\begin{equation}\label{covmatstatic}
    \dv{}{t}\bar\sigma^j = \bar{A}_{j}\bar\sigma^j+\bar\sigma^j\bar{A}_{j}^T+\bar{C}_j,
\end{equation}
where we have defined the ladder operator vector as $\hat{\bm{B}}_j \equiv [\hat{b}_{ji},\hat{b}_{ji}^\dagger,\hat{b}_{jp},\hat{b}_{jp}^\dagger]^T$ and its  covariance matrix as
\begin{align}
    \bar\sigma^j_{\lambda\mu} \equiv \frac{1}{2}  \langle\hat{B}_{j\lambda}\hat{B}_{j\mu} + \hat{B}_{j\mu}\hat{B}_{j\lambda}\rangle -\langle\hat{B}_{j\lambda}\rangle\langle\hat{B}_{j\mu}\rangle,
\end{align}
for $\lambda,\mu=1,2,3,4$. The $4\times4$ drift matrix $\bar{A}_j$ is
\begin{multline}
    \bar{A}_j\equiv \text{diag}[-i\Omega_{ji}',i\Omega_{ji}',-i\Omega_{jp}',i\Omega_{jp}']+
    \\+\left[\begin{array}{c|c}
       -(\gamma^{\rm dop}/2)\mathbb{1}_{2\times 2}  & -ig_j \mathbb{Z} \mathbb{J}_{2\times 2} \\ \hline
        -ig_j \mathbb{Z} \mathbb{J}_{2\times 2}  & -((\gamma^{\rm fb}+\gamma^{\rm gas})/2)\mathbb{1}_{2\times 2}
    \end{array}\right],
\end{multline}
where $\mathbb{Z}$ is the Pauli-Z matrix and $\mathbb{J}_{n\times n}$ is a $n\times n$ matrix whose entries are all equal to $1$. 
The $4\times4$ diffusion matrix $\bar{C}_j$ is
\begin{align}
    \bar{C}_j \equiv \mqty[
    0 && \Gamma_j^{\rm dop} && - i g_j && 0 \\
    \Gamma_j^{\rm dop} && 0 && 0 && i g_j \\
    -i g_j && 0 && \Gamma_j^{\rm gas} - \Gamma_{jp} && \Gamma_{jp} \\
    0 && i g_j && \Gamma_{jp} && \Gamma_j^{\rm gas} - \Gamma_{jp}
    ],
\end{align}
where we have defined the total nanoparticle heating rate, $\Gamma_{jp}\equiv\Gamma_j^{\rm gas}+\Gamma^{\rm td}_j+\Gamma_j^{\rm ba}$. From the covariance matrix $\bar\sigma^j$, we extract the steady-state phonon occupation of the nanoparticle along each axis $j=x,y,z$, i.e., $\langle \hat n_{jp}\rangle_{\rm ss}\equiv\langle \hat b_{jp}^\dagger\hat b_{jp}\rangle_{\rm ss}$. The full analytical expression for this phonon occupation number is given in Appendix~\ref{Appendixformula}.

The nanoparticle steady-state phonon occupation numbers along the $z$ axis and the $x$ axis are shown in Fig.~\ref{fig:n-vs-gammap,Gammap} for a continuously cooled ion and for the parameters in Table~\ref{tab:parameters}, as a function of the total nanoparticle damping rate, $\gamma_p\equiv\gamma^{\rm fb}+\gamma^{\rm gas}$. For this figure we assume $\Gamma_j^{\rm td}=0$, but the effect of these additional noise sources is discussed below. The dotted black line in in Fig.~\ref{fig:n-vs-gammap,Gammap} shows the case where ion and particle are not coupled ($g_j=0$) and thus no sympathetic cooling is possible. The blue and green dotted lines mark two relevant values of damping, namely the damping in the absence of optical feedback ($\gamma_p = \gamma^{\rm gas}=2\pi\times 8.4$ nHz) and the typical damping values achieved with feedback cooling in state-of-the-art experiments~\cite{Bykov_2025} ($\gamma_p \approx \gamma^{\rm fb}=2\pi\times 1$ Hz). 
In general, the nanoparticle occupation numbers in Fig.~\ref{fig:n-vs-gammap,Gammap} are always lower than their room-temperature values (
which are $5.9\times10^{9}$ in the $z$ axis and 
$4.51\times10^{9}$ in the $x$ axis
, respectively), indicating motional cooling for all values of $\gamma_p$ across two distinct regimes. First is the sympathetic cooling regime for $\gamma_p \ll \gamma_p^{\rm eff}$, where  
the nanoparticle is cooled by the ion via Coulomb interaction at a rate $\gamma_p^{\rm eff}$ defined below. In this regime, assuming  weak ion-nanoparticle coupling and ion mechanical damping, i.e., $\vert g_j\vert, \gamma^{\rm dop} \ll \Omega_{ji}'$ (which is fulfilled for the parameters of Table~\ref{tab:parameters}; see Fig.~\ref{fig:g,D-vs-Qp,Mp}), and neglecting the nanoparticle damping $\gamma_p$, a simple analytical approximation for the nanoparticle occupation number can be derived:
\begin{equation}\label{Npanalyticalapprox}
    \expval{\hat{n}_{jp}}_{\rm ss} \approx \frac{1}{\gamma^{\rm dop}} \frac{1}{g_j^2} \frac{ \left(\Omega_{ji}^{'2} - \Omega_{jp}^{'2}\right)^2}{4 \Omega_{ji}^{'} \Omega_{jp}^{'}} \Gamma_{jp}.
\end{equation}
The above expression, indicated by a horizontal dashed line in Fig.~\ref{fig:n-vs-gammap,Gammap}a,  highlights the main limitation to sympathetic cooling, namely, the vastly different mechanical frequencies of ion and nanoparticle, $\Omega_{ji}^{'2} \gg  \Omega_{jp}^{'2}$. Despite this mismatch, Fig.~\ref{fig:n-vs-gammap,Gammap} shows that in the absence of optical feedback sympathetic cooling from a single ion can reduce nanoparticle occupation along the $z$ axis by three orders of magnitude. From equation~\eqref{Npanalyticalapprox} we 
can define the sympathetic cooling rate as $\gamma_{p}^{\rm eff} \equiv \Gamma_{jp}/\expval{\hat{n}_{jp}}_{\rm ss}$, which, in the limit $\Omega_{ji}^{'2} \gg  \Omega_{jp}^{'2}$, is
\begin{equation}\label{gammaDOP}
    \gamma_{p}^{\rm eff} \approx \gamma^{\rm dop}  \frac{4 \Omega_{jp}^{'}}{ \Omega_{ji}^{'3}} g_j^2 .
\end{equation}

The second cooling regime in Fig.~\ref{fig:n-vs-gammap,Gammap} is the feedback cooling regime $\gamma_p^{\rm eff}\lesssim \gamma^{\rm fb}$, where the damping is dominated by the feedback contribution and the steady-state occupation can be written as $\langle\hat n_{jp}\rangle_{\rm ss}\approx \Gamma_{jp}/\gamma^{\rm fb}$. This expression describes an initial linear decrease with $\gamma_p \approx \gamma^{\rm fb}$, followed by a saturation to a constant $\langle\hat n_{jp}\rangle_{\rm ss}\to \Gamma_j^{\rm ba}/\gamma^{\rm fb}$ when the measurement backaction rate given in Eq.~\eqref{Gammarecoil} becomes dominant over other heating rates. We remark that both the onset of this saturation and the minimum achievable occupation number depend strongly on the experimental measurement setup. The results in Fig.~\ref{fig:n-vs-gammap,Gammap} correspond to the measurement scheme in Ref.~\cite{Bykov_2025}, namely, $\varepsilon=2.11$, $\lambda_0= 780~\mathrm{nm}$, and $c_{\rm fb}= 1.57\times10^{-6}~\mathrm{Hz~m^2/W}$, and $\zeta=7$ \footnote{Note that in regards to experiment Ref.~\cite{Bykov_2025}, this choice of $\zeta$ leads to an overestimation of the heating rate, as it assumes a probe beam propagating exactly along the motional axis.}.

In the current experiments~\cite{Bykov_2025}, nanoparticle heating is dominated by trap displacement-like noise with rates $\Gamma_z^{\rm td}=2\pi\times6.3~\mathrm{kHz}$ and $\Gamma_x^{\rm td}=2\pi\times4.9~\mathrm{kHz}$. This leads to phonon occupation numbers about 2 orders of magnitude larger than those without trap displacement noise, as shown in the inset of panel (a). In this case, the ion is still able to significantly cool the nanoparticle, sympathetically reducing its room-temperature occupation also by three orders of magnitude along the $z$ direction. In this direction, our model predicts steady-state nanoparticle temperatures $T\approx 22~\mathrm{K}$ in current experiments. In principle, the additional noise sources responsible for $\Gamma_j^{\rm td}$ can be suppressed by e.g. reducing the electrical noise and improving vibrational isolation. If such noise is fully suppressed, our model predicts that steady-state nanoparticle temperatures can reach $T\approx 0.17~\mathrm{K}$ via sympathetic cooling with a single ion at room temperature. For deeper cooling toward the ground state, however, measurement-based feedback remains the most effective route.

\begin{figure}[h]
    \centering
    \includegraphics[width=\linewidth]{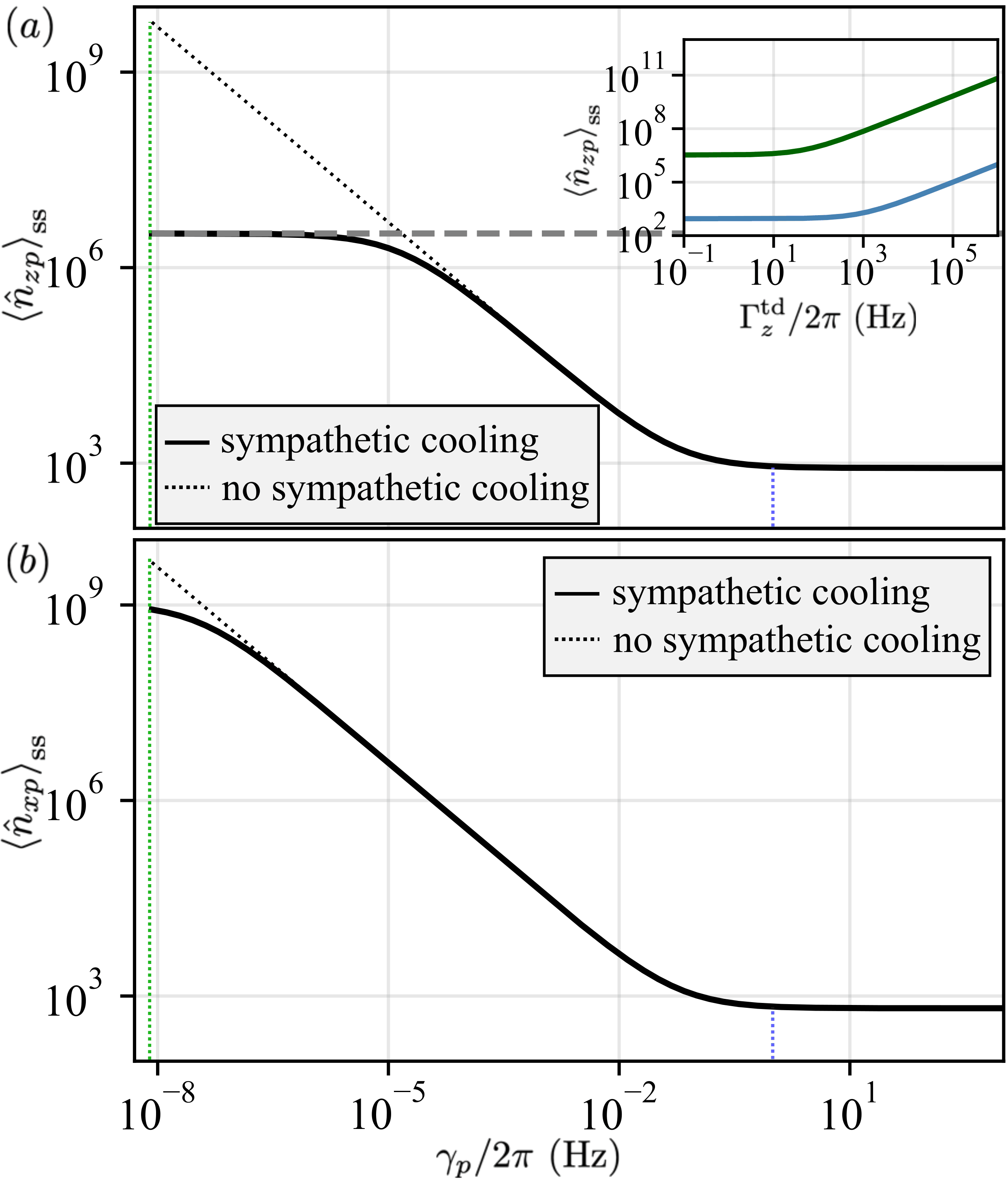}
    \caption{Steady-state phonon number of nanoparticle motion along the (a) $z$ axis and (b) $x$ axis versus nanoparticle damping rate, $\gamma_p = \gamma^{\rm fb}+\gamma^{\rm gas}$, without additional trap noise, $\Gamma_j^{\rm td}=0$, and with the ion being continuously Doppler cooled. Parameter values are taken from Table~\ref{tab:parameters}. The dotted black curves correspond to having no sympathetic cooling, i.e. $g_j=0$, and can be well approximated by $\langle\hat n_{jp}\rangle_{\rm ss}\approx \Gamma_{jp}/\gamma_{p}$ for $\gamma_p\lesssim2\pi\cdot 10^{-1}~\rm{Hz}$. The horizontal gray dashed line in (a) shows our approximate expression for the occupation in the sympathetic cooling regime, Eq.~\eqref{gammaDOP}. The inset in (a) shows the steady-state phonon number as a function of the trap-displacement noise in the system, $\Gamma^{\rm td}_z$, for $\gamma_p = 2 \pi \times 8.4 ~\mathrm{nHz}$ (green) and $\gamma_p = 2 \pi \times 1 ~\mathrm{Hz}$ (blue). These two values of damping are marked in the main panels by dotted green and blue lines, respectively.}
    \label{fig:n-vs-gammap,Gammap}
\end{figure}

\subsection{Effect of micromotion}\label{sec:micromotion}

When the equilibrium position of the trapped objects is not at the center of the linear Paul trap, the secular approximation to the ion and nanoparticle Hamiltonians, Eq.~\eqref{Hsigmasecular}, can become inaccurate due to micromotion, i.e., due to the 
terms oscillating at the RF frequencies in Eq.~\eqref{rclassicalDehmelt} becoming relevant. In this section we quantify the effect of micromotion on sympathetic cooling by solving the master equation, Eq.~\eqref{eq:basic-master-eq}, under substitution of the Hamiltonian Eq.~\eqref{Hsigmasecular} by the full time-dependent version
\begin{equation}\label{fullHtimedep}
   \hat{H}_{\sigma}(t) = \frac{\hat{\mathbf{P}}^2_\sigma}{2M_\sigma}+\frac{M_\sigma}{2}\sum_{j=x,y,z}W_{ j\sigma}(t)\hat{R}^2_{ j\sigma}, 
\end{equation}
with the function $W_{ j\sigma}(t)$ given in Eq.~\eqref{Wpotentialdef}. We linearise the Coulomb interaction around the static equilibrium positions, i.e., around the same equilibrium positions $\bm{D}_i$ and $\bm{D}_p$ determined only from the secular Hamiltonian and derived in Sec.~\ref{sec:coulomb-interaction}. The linearised Hamiltonian around these static equilibrium positions is
\begin{multline}\label{HipquantumMicromotion}
    \hat{H}_{\rm{tot}}(t)= 
          \frac{1}{2}\hat{\bm{P}}^T \bar{M}  \hat{\bm{P}} + \frac{1}{2} \hat{\bm{X}}^T  \bar{V}_m(t) \hat{\bm{X}}+ \mathbf{F}(t)\hat{\bm{X}},
\end{multline}
where the position and momenta vectors $\hat{\bm{X}}$ and $\hat{\bm{P}}$ are defined in Eqs.~(\ref{Xdef}-\ref{Pdef}), the matrix $\bar{M}$ is defined in Eq.~\eqref{Massmat}, and the new generalised potential matrix takes a similar  form
as Eq.~\eqref{Vatrixdef} but with time-dependent frequencies,
\begin{equation}\label{Vmatrixmicromotion}
    \bar{V}_m(t) \equiv 
    \left[\begin{array}{cc}
        -\bar{N}_{\rm coul} & \bar{N}_{\rm coul} \\
         \bar{N}_{\rm coul}& -\bar{N}_{\rm coul}
    \end{array}\right]+\!\left[\begin{array}{cc}
        \bar{V}_{di,m}(t) &\!
        \mathbb{0}_{3\times 3} \\
        \mathbb{0}_{3\times 3} & 
        \! \! \!\! \bar{V}_{dp,m}(t)\!\!
    \end{array}\right]\!,
\end{equation}
with $\bar{V}_{di,m}(t)\equiv M_i \text{diag} [W_{xi}^2(t),W_{yi}^2(t),W_{zi}^2(t)]$ and $\bar{V}_{dp,m}(t)\equiv M_p\text{diag}[W_{xp}^2(t),W_{yp}^2(t),W_{zp}^2(t)]$. Equation~\eqref{HipquantumMicromotion} also contains a time-dependent force describing the driving of the mean oscillator position due to the RF fields. This vector can be written as $\mathbf{F}(t)=[\mathbf{F}_i(t),\mathbf{F}_p(t)]^T$
with
\begin{equation}
\mathbf{F}_\sigma(t)=M_\sigma\left[
\begin{array}{c}
     (W_{x\sigma}(t)-\Omega_{x\sigma}^2)d_{x\sigma}  \\
     (W_{y\sigma}(t)-\Omega_{y\sigma}^2)d_{y\sigma}
     \\
    (W_{z\sigma}(t)-\Omega_{z\sigma}^2)d_{z\sigma} 
\end{array}
\right] .
\end{equation}
Equation~\eqref{HipquantumMicromotion} is an accurate description of the system provided that each trapped particle remains close to its equilibrium position even in the presence of micromotion, i.e., $\langle \delta \hat R_{j\sigma}\delta \hat R_{j'\sigma'}\rangle (t) \ll D^2$. We assume that the dissipators $\mathcal{D}_p \left( \hat{\rho} \right)$ and $\mathcal{D}_i \left( \hat{\rho} \right)$ in Eq.~\eqref{eq:basic-master-eq} are unaffected by micromotion, but, since the ladder operators $\hat{b}_{j \sigma}$ are not well defined for time-dependent Hamiltonians, we write these dissipators in terms of position and momentum operators $\delta\hat R_{j\sigma}$ and $\hat P_{j\sigma}$ using the identity $2\hat{b}_{j \sigma}=(\delta \hat R_{j\sigma}/R_{j\sigma}^{\rm zpf})+i(\hat{P}_{j\sigma}/P_{j\sigma}^{\rm zpf})$.

\begin{figure*}[t]
    \centering
    \includegraphics[width=\linewidth]{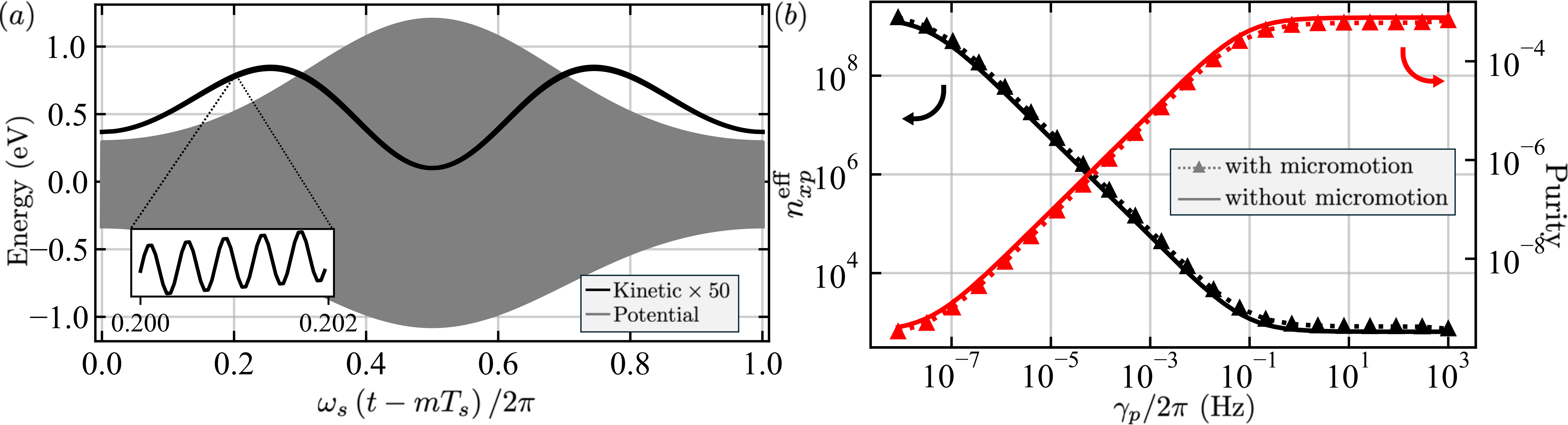}
    \caption{(a) Potential and kinetic energy of nanoparticle motion along the $x$ axis as a function of time in the long-time limit. The kinetic energy is multiplied by a factor 50 for visibility. The inset shows the fast oscillations of the micromotion at frequency $\omega_f$. (b) Time-averaged purity (Eq.~\eqref{averagepurity}) and effective occupation number (Eq.~\eqref{neff}) of the nanoparticle motion along the $x$ axis in the long-time limit as a function of nanoparticle damping $\gamma_p=\gamma^{\rm gas}+\gamma^{\rm fb}$. Triangles and solid lines indicate the values with and without micromotion. Parameters are taken from Table~\ref{tab:parameters}.}
    \label{fig:energy-vs-time-and-n,purity-vs-gammap}
\end{figure*}

In an analogous way to the previous subsection, but this time using the time-dependent Hamiltonian~\eqref{HipquantumMicromotion}, we derive the equations of motion
\begin{equation}\label{eomsingleV}
    \dv{}{t}\langle \hat{\mathbf{v}}\rangle =\bar{A}_m(t) \langle \hat{\mathbf{v}}\rangle -\left[\begin{array}{c}
         \mathbb{0}_{6\times 1}  \\
         \mathbf{F}(t) 
    \end{array}\right],
\end{equation}
and 
\begin{equation}\label{eomcovmat}
    \dv{}{t} \bar\Sigma_{\rm c}=\bar{A}_m(t)\bar\Sigma_{\rm c}+\bar\Sigma_{\rm c}\bar{A}_m^T(t)+\bar{C}_m,
\end{equation}
where we define the $12-$dimensional quadrature vector $\hat{\mathbf{v}}\equiv[\hat{\mathbf{X}}, \hat{\mathbf{P}}]^T$ and the $12\times 12$ covariance matrix of the quadratures $(\bar\Sigma_{\rm c})_{kl}\equiv \langle\hat{v}_k\hat{v}_l+\hat{v}_l\hat{v}_k\rangle/2-\langle\hat{v}_k\rangle\langle\hat{v}_l\rangle$, and 
where
\begin{equation}
    \bar{A}(t)\equiv \left[\begin{array}{cc}
        -\bar{\gamma}/2 & \bar{M} \\
        -\bar{V}_m(t) &  -\bar{\gamma}/2
    \end{array}\right],
\end{equation}
with
\begin{equation}
    \bar\gamma\equiv\left[\begin{array}{cc}
        \gamma^{\rm dop}\mathbb{1}_{3\times 3} & \mathbb{0}_{3\times 3} \\
        \mathbb{0}_{3\times 3} & (\gamma^{\rm fb}+\gamma^{\rm gas})\mathbb{1}_{3\times 3}
    \end{array}\right].
\end{equation}
The inhomogeneous matrix term in Eq.~\eqref{eomcovmat} is given by 
\begin{equation}
    \bar{C} = \mqty[
    \bar{C}_{rr} && \mathbb{0}_{6\times6} \\
    \mathbb{0}_{6\times6} && \bar{C}_{pp}
    ],
\end{equation}
where the diagonal blocks are given by
\begin{align}
\begin{split}
    \bar{C}_{rr} = 
    \mathrm{diag} \Big[
    &(R^{\text{zpf }}_{xi})^2 \gamma^{\rm dop},
    (R^{\text{zpf }}_{yi})^2 \gamma^{\rm dop},
    (R^{\text{zpf }}_{zi})^2 \gamma^{\rm dop},\\
    &(R^{\text{zpf }}_{xp})^2 \gamma^{\rm gas},
    (R^{\text{zpf }}_{yp})^2 \gamma^{\rm gas},
    (R^{\text{zpf }}_{zp})^2 \gamma^{\rm gas}
    \Big] + \\
    + 2 \mathrm{diag} \Big[  
    &(R^{\text{zpf }}_{xi})^2 \Gamma_x^{\rm dop}, 
    (R^{\text{zpf }}_{yi})^2 \Gamma_y^{\rm dop},(R^{\text{zpf }}_{zi})^2 \Gamma_z^{\rm dop},\\
    &(R^{\text{zpf }}_{xp})^2 \Gamma_x^{\rm gas}, (R^{\text{zpf }}_{yp})^2 \Gamma_y^{\rm gas},(R^{\text{zpf }}_{zp})^2 \Gamma_z^{\rm gas}
    \Big],
\end{split}
\end{align}
and a similar expression for $\bar{C}_{pp}$ under the substitutions $R^{\rm zpf}_{j\sigma}\to P^{\rm zpf}_{j\sigma}$ and $\Gamma^{\rm gas}_j\to 2\Gamma_{jp} - \Gamma^{\rm gas}_j$.

By vectorising the covariance matrix we cast both dynamical Eqs.~(\ref{eomsingleV}-\ref{eomcovmat}) as a Floquet linear system of differential equations of the form $\dv*{\mathbf{f}}{t}=\bar{B}(t)\mathbf{f}+\bm{w}(t)$ where $\bar{B}(t)$ is, assuming integer $l^{-1}$,  a periodic matrix with period $T_s=2\pi/\omega_s$ (for the parameters of Table~\ref{tab:parameters}, $l^{-1}=2500$). 
According to Floquet theory \cite{Andreevich_1975, Slane_2019}, the solution at times $m T_s+t$ with $t<T_s$, $m$ integer, and $m\to\infty$ is also periodic with period $T_s$, and reads
\begin{multline}\label{Floquetsolution}
    \mathbf{f}(m T_s+t)=\bar{X}(t)\Bigl[\left(\mathbb{1} - \bar{X}(T_s)\right)^{-1} \bar{X}(T_s) \bm{\Lambda}(T_s)  \\+ \bm{\Lambda}(t) \Bigr],
\end{multline}
with 
\begin{equation}
    \bm{\Lambda}(t)\equiv \int_0^t \bar{X} (\tau)^{-1} \bm{w} (\tau) \dd\tau.
\end{equation}
Here $\bar{X}(t)$ is the fundamental matrix solution of the homogeneous system of equations, namely the solution of $\dv*{\bar{X}(t)}{t}=\bar{B}(t)\bar{X}(t)$ with $\bar{X}(0)=\mathbb{1}$. The above expressions reduce the solution of the system of equations to the computation of $\bar{X}(t)$ across one time period only.
The solution Eq.~\eqref{Floquetsolution} is dynamically stable provided that the eigenvalues of the matrix $\bar{X}(T_s)$, denoted by $\lambda_k$, lie within the unit circle in the complex plane, i.e.,
\begin{align}\label{Floquetstability}
    \max_k{\left( \abs{\lambda_k} \right)} \leq 1.
\end{align}
This is the Floquet stability criterion employed in Sec.~\ref{sec:coulomb-interaction}. We will employ it again in Sec.~\ref{sec:results-N}.

Let us focus on the specific case study in this work, where the equilibrium positions of ion and nanoparticle lie along the $z$-axis (i.e. $d_{x\sigma}=d_{y\sigma}=0$), and only a DC voltage is applied to the endcap electrodes, i.e. $W_{z\sigma}(t)=\Omega_{z\sigma}^2$. In this case $\mathbf{F}(t)=0$ and, in analogy with the previous section, the dynamics also decouple into three subsets of two coupled harmonic oscillators, each corresponding to ion and nanoparticle motion along one Cartesian axis. In optimized traps (see Table~\ref{tab:parameters}) the motion along $z$ is only governed by the DC potential and thus the dynamics remain time-independent. In other words, only the motional degrees of freedom along the $x$ and $y$ axes are affected by micromotion. Since the dynamics along these axes are qualitatively the same, hereafter we focus on the motion along the $x$ axis. 

We compute the Floquet solution Eq.~\eqref{Floquetsolution} using the built-in Dormand-Price solver in Python. The obtained
kinetic and potential energies of the nanoparticle motion are plotted in Fig.~\ref{fig:energy-vs-time-and-n,purity-vs-gammap}(a). The highly oscillatory behavior is due to the dual-frequency nature of the trap, whose two periods $T_s$ and $T_f\equiv 2\pi/\omega_f \ll T_s$ are inherited by the Floquet solutions. Accurately recovering this fine structure requires a fine time resolution (see inset). 
Using the Floquet solution, we compute the motional purity of the nanoparticle averaged over one period,
\begin{align}\label{averagepurity}
    \mu_x\equiv \frac{1}{T_s}\int_0^{T_s}  \dd t \frac{\hbar}{2\sqrt{\det(\bar{\Sigma}_{xp}(t))}},
\end{align}
where $\bar{\Sigma}_{jp}$ is the nanoparticle covariance matrix, with $(\bar{\Sigma}_{jp})_{11}=\langle \delta\hat R_{jp}^2\rangle-\langle \delta\hat R_{jp}\rangle^2$, $(\bar{\Sigma}_{jp})_{22}=\langle \hat P_{jp}^2\rangle-\langle \hat P_{jp}\rangle^2$, and $(\bar{\Sigma}_{jp})_{12}=(\bar{\Sigma}_{jp})_{21}=\langle\{\delta\hat R_{jp}, \hat P_{jp}\}\rangle/2- \langle \delta\hat R_{jp}\rangle\langle \hat P_{jp}\rangle$. 
The above expression for the purity is valid for Gaussian states, which always remain Gaussian under quadratic time evolution~\cite{Barthel_2022}. From this average purity we define an effective occupation number through the identity~\cite{Paris_2003}
\begin{align}\label{neff}
     n_{xp}^{\rm eff} &\equiv \frac{1}{2}\left(\frac{1}{\mu_x} - 1\right).
\end{align}
In the absence of micromotion the above effective occupation number becomes the true occupation defined in Sec.~\ref{sec:sympathetic-cooling}.
Both the average purity and the effective occupation are displayed in Fig.~\ref{fig:energy-vs-time-and-n,purity-vs-gammap}(b) as a function of the nanoparticle damping rate, $\gamma_p$, for the same parameters as used in Fig.~\ref{fig:n-vs-gammap,Gammap}. Micromotion has a small effect on the cooling, increasing the effective occupation by 15-25\%.

\section{Sympathetic cooling with multiple ions}\label{sec:sympathetic-cooling-N}

State-of-the-art electrical traps are capable of co-trapping more than one ion along with the nanoparticle \cite{Bykov_2025}. Since all the ions can be individually Doppler-cooled, each can act as an energy sink, thus enhancing the performance of sympathetic cooling. In this section, we quantify the scaling of sympathetic cooling with the number of ions. First, we summarize how to generalise our theory to $N$ ions for arbitrary trap parameters in Sec.~\ref{sec:hamiltonian-master-equation-N}. Then, in Sec.~\ref{sec:results-N}, we show the cooling performance of the $N-$ion ensemble for the specific parameters of Table~\ref{tab:parameters}.

\subsection{Hamiltonian and master equation}\label{sec:hamiltonian-master-equation-N}

To analyze the coupled dynamics of a nanoparticle and $N>1$ identical ions, we generalise the approach employed in Sec.~\ref{sec:coupled-quantum-dynamics}. We focus on the secular dynamics since, as we will see below, for the parameters considered in this work the micromotion does not affect the motional cooling of the nanoparticle along the $z$-axis. As a first step, we generalise the Coulomb interaction Eq.~\eqref{Coulombclassical} to multiple ions as
\begin{align}\label{eq:coulomb-classical-N}
    V_{ip} &= \frac{Q_i}{4 \pi \varepsilon_0}\Bigg[\sum_{k=1}^N  \frac{ Q_p}{\abs{\bm{R}_{i,k} - \bm{R}_p }} + \sum_{\substack{k=1\\l>k}}^N  \frac{Q_i}{\abs{\bm{R}_{i,k} - \bm{R}_{i,l} }}\Bigg].
\end{align}
Note that this expression also includes the all-to-all interactions between the ions.
The equilibrium positions of all the objects in the trap, $\{\bm{D}_{i,1},...,\bm{D}_{i,N},\bm{D}_p\}$, are given by the same zero-force condition, Eq.~\eqref{eq:equilibrium-config-condition}, which in this case becomes a system of $3N+3$ nonlinear equations,
\begin{equation}\label{eq:equilibrium-config-conditionNpart}
    \left(M_p\Omega_{jp}^2 R_{jp}+\frac{\partial}{\partial R_{ jp}}V_{ip}\right)_{\mathbf{R}_{i,j}=\mathbf{D}_{i,j},\mathbf{R}_p=\mathbf{D}_p}=0,
\end{equation}
\begin{equation}\label{eq:equilibrium-config-conditionNions}
    \left(M_i\Omega_{ji}^2 R_{ji,k}+\frac{\partial}{\partial R_{ ji,k}}V_{ip}\right)_{\mathbf{R}_{i,j}=\mathbf{D}_{i,j},\mathbf{R}_p=\mathbf{D}_p}=0,
\end{equation}
for $j=x,y,z$ and $k=1,...,N$.
We define displacements from the equilibrium positions as in Eq.~\eqref{dispfromeq}, namely $\hat{\mathbf{R}}_{p}=\bm{D}_p+\delta\hat{\mathbf{R}}_{p}$ with 
$\delta\hat{\mathbf{R}}_{p} = [\delta\hat R_{xp},\delta\hat R_{yp},\delta\hat R_{zp}]^T $, and $\hat{\mathbf{R}}_{ji,k}=\bm{D}_{i,k}+\delta\hat{\mathbf{R}}_{ji,k}$ with 
$\delta\hat{\mathbf{R}}_{i,k} = [\delta\hat R_{xi,k},\delta\hat R_{yi,k},\delta\hat R_{zi,k}]^T $, and linearize the Coulomb interaction Eq.~\eqref{eq:coulomb-classical-N} around this  equilibrium configuration by expanding it 
to second order in the variables $\delta\hat R_{jp}/\vert \bm{D}_p\vert$ and $\delta\hat R_{ji,k}/\vert \bm{D}_{i,k}\vert$. The total linearised Hamiltonian can be written as
\begin{multline}\label{eq:hamiltonian-quantum-N }
    \hat{H}_{\rm tot} \equiv \hat{H}_p + \sum_{k=1}^N \hat{H}_{i,k} + \hat{V}_{ip} =\\= 
    \frac{1}{2}\hat{\bm{P}}^T \bar{M}_N  \hat{\bm{P}} + \frac{1}{2} \hat{\bm{X}}^T  \bar{V}_N \hat{\bm{X}},
\end{multline}
where the individual Hamiltonians of the nanoparticle, $\hat{H}_p$, and of each ion, $\hat{H}_{i,k}$, are given by Eq.~\eqref{Hsigmasecular}, and where the position and momentum operator vectors are generalised as $\hat{\bm{X}}=[\delta\hat {\mathbf{R}}_{i,1},\delta\hat {\mathbf{R}}_{i,2},...,\delta\hat {\mathbf{R}}_{i,N},\delta\hat {\mathbf{R}}_{p} ]^T$ and $\hat{\bm{P}}=[\hat {\mathbf{P}}_{i,1},\hat {\mathbf{P}}_{i,2},...,\hat {\mathbf{P}}_{i,N},\hat {\mathbf{P}}_{p} ]^T$.
The inverse mass matrix and generalised potential matrix, $\bar{M}_N$ and $\bar{V}_N$, have dimensions $(3N+3) \times (3N+3)$ and are 
\begin{equation}\label{MassmatN}
    \bar{M}_N \equiv 
    \left[\begin{array}{c|c}
        M_i^{-1}\mathbb{1}_{3N\times 3N} & \mathbb{0}_{3N\times 3} \\
        \hline \mathbb{0}_{3\times 3N}
         & M_p^{-1}\mathbb{1}_{3\times 3}
    \end{array}\right]
\end{equation}
and 
\begin{widetext}
\begin{equation}\label{VatrixdefNions}
    \bar{V}_N \equiv 
    \left[\begin{array}{ccccc}
        -\bar{N}_{1} - \bar{N}_{1p} + \bar{V}_{di} & \bar{N}_{12}& \bar{N}_{13}&\dots & \bar{N}_{1p} \\
        \bar{N}_{21} & -\bar{N}_{2} - \bar{N}_{2p}+\bar{V}_{di} & \bar{N}_{23} & \dots &\bar{N}_{2p}
        \\
        \bar{N}_{31} & \bar{N}_{32} & -\bar{N}_{3} - \bar{N}_{3p} + \bar{V}_{di} & \dots & \bar{N}_{3p}
        \\
        \vdots & \vdots &\vdots &\ddots &  \vdots \\
        \bar{N}_{p1} & \bar{N}_{p2} & \bar{N}_{p3} & \dots & -\sum_{k=1}^N \bar{N}_{kp} + \bar{V}_{dp}
    \end{array}\right].
\end{equation}
\end{widetext}
Here we have defined the generalised Coulomb matrix (see Eq.~\eqref{Ncoulomb})
\begin{multline}
   \bar{N}_{kl} \equiv  \frac{Q_i^2}{4 \pi \varepsilon_0\vert \bm{D}_{i,k}-\bm{D}_{i,l}\vert^3} \times\\\times \left[\mathbb{1}_{3\times 3}-3\frac{(\bm{D}_{i,k}-\bm{D}_{i,l})\!\otimes\!(\bm{D}_{i,k}-\bm{D}_{i,l})}{\vert \bm{D}_{i,k}-\bm{D}_{i,l}\vert^2}\right],
\end{multline}
with $k,l=1,2,...,N$, and with $\bar{N}_{k}=\sum_{l=1,l\ne k}^N\bar{N}_{kl}$. The matrices $\bar{N}_{kp}=\bar{N}_{pk}$ have an analogous expression under the substitutions $Q_i^2\to Q_i Q_p$ and $\bm{D}_{i,l}\to \bm{D}_{p}$.

The Hamiltonian in the form Eq.~\eqref{eq:hamiltonian-quantum-N } can be introduced into the secular master equation Eq.~(\ref{eq:basic-master-eq}), which for the $N-$ion case is
\begin{align}\label{eq:basic-master-eqNions}
    \dot{\hat{\rho}} &= -\frac{i}{\hbar} \comm{\hat{H}_{\rm tot}}{\hat{\rho}} + \mathcal{D}_p \left( \hat{\rho} \right) + \sum_{k=1}^N\mathcal{D}_{i,k} \left( \hat{\rho} \right).
\end{align}
Here $\hat\rho$ is the density matrix of the motional degrees of freedom of the $N$ ions and the nanoparticle, and $\mathcal{D}_p$ and each of the ion dissipators $\mathcal{D}_{i,k}$ are given by Eqs.~(\ref{dissipatorP}) and Eq.~(\ref{dissipatorI}) respectively.
From the above master equation the steady-state occupation of each degree of freedom can be extracted as detailed in Sec.~\ref{sec:sympathetic-cooling}.

\subsection{Results}\label{sec:results-N}

Let us focus on the results obtained for the parameters of Table~\ref{tab:parameters}. First, we obtain the stable equilibrium positions by solving the nonlinear system of Eqs.~(\ref{eq:equilibrium-config-conditionNpart}-\ref{eq:equilibrium-config-conditionNions}) for each value of $N$ using SciPy's root-finding function. We run the solver $10^6$ times, each with different seed values for the $3N+3$ spatial coordinates of all the objects. At each run, the seed values are drawn randomly within a cubic box of size $200~\mathrm{\mu m}\times200~\mathrm{\mu m}\times200~\mathrm{\mu m}$ centered at the origin. From each of the equilibrium configurations found, we retain only those which fulfill \textit{both} the dynamical and the Floquet stability criteria, defined respectively in Secs.~\ref{sec:coulomb-interaction} and ~\ref{sec:micromotion} (See Eqs. \eqref{Kdef} and~\eqref{Floquetstability}). Similarly to the $N=1$ case, all the stable configurations correspond to the equilibrium positions of all the objects -- nanoparticle and all $N$ ions -- lying along the $z$-axis.
We ascribe this to the fact that this configuration minimises the negative Coulomb-induced shift in the mechanical frequencies along the $z$-axis. More specifically, as shown by Eq.~\eqref{frequencyrenorm}, due to the Coulomb interaction the trap frequencies of any two objects are always reduced in the directions orthogonal to the line connecting them. If this shift is large enough the mechanical frequencies can approach zero or even become imaginary, corresponding to  unstable dynamics.
When all objects lie along the $z$-axis only the radial trapping frequencies $\{\Omega_{xp},\Omega_{yp},\Omega_{xi,k},\Omega_{yi,k}\}$ decrease, whereas for any other configuration the longitudinal frequencies $\Omega_{zp}$ and $\Omega_{zi,k}$ also do. Since $\Omega_{zp}<\Omega_{xp},\Omega_{yp}$ and $\Omega_{zi,k}<\Omega_{xi,k},\Omega_{yi,k}$ (see Table~\ref{tab:parameters}), the Coulomb-induced shift has a stronger de-stabilising effect in the latter case as it reduces frequencies that are already small. Hence, configurations where all objects lie along the $z$-axis tend to be more stable. For each value of $N$ multiple stable equilibrium configurations exist, but they result in very similar values for $\langle \hat n_{zp}\rangle_{\rm ss}$ with deviations of $\sim0.01\%$.

No stable solutions were found for $N>8$, suggesting that $N=8$ is the 
maximum number of ions that can be co-trapped along a single axis for the chosen parameter values~\footnote{For $N=9$ an additional stable solution was found appeared in less than $0.01$\% of the runs. This solution was discarded as it corresponds to a weakly stable minimum occupying a very small volume in coordinate space, and thus unlikely to confine the objects under the influence of small -- e.g. thermal -- fluctuations.}. Note that this number is below the prediction for the number of ions one can stably trap in a single-frequency trap ($N \approx 13$ for our parameters according to Ref.~\cite{Enzer_2000}). Physically, this is expected as the nanoparticle has large mass and charge and thus pushes the ions away from it, leading to smaller ion-ion separations along $z$ and thus to the onset of Coulomb-induced instability at lower $N$. 
We remark that this is not a fundamental limitation but rather a constraint imposed by the choice of parameters in Table~\ref{tab:parameters}. For instance, increasing the trap stiffness in the radial directions could enable trapping more ions in this configuration. 
We do not explore such improvements in this work as $N\le 8$ is sufficient for the purposes of this section, namely to quantify the scaling of sympathetic cooling performance with $N$.

Since the equilibrium positions lie on the $z$-axis, the motional degrees of freedom along each axis decouple. We thus focus hereafter only on the dynamics of each object along the $z$-axis, which is governed by the $(N+1)\times(N+1)$ Hamiltonian (see Eq.~\eqref{eq:hamiltonian-quantum-N })
\begin{multline}\label{HN_onlyz}
    \hat{H}_{\rm tot}^{(z)} = \hat{H}_{p}^{(z)} +\sum_{k=1}^N \hat{H}_{i,k}^{(z)} + \sum_{k=1}^N\delta{\hat{R}}_{zi,k}
    \\ \times\Big(\big(\bar{N}_{kp}\big)_{33}\delta{\hat{R}}_{zp}+\sum_{l>k}
    \big(\bar{N}_{kl}\big)_{33}\delta{\hat{R}}_{zi,l}
    \Big)
    ,
\end{multline}
with 
\begin{equation}
    \hat{H}_{p}^{(z)} = \frac{\hat{P}_{zp}^2}{2M_p} + \frac{M_p}{2}\Omega_{zp}^{'2}\delta{\hat{R}}_{zp}^2,
\end{equation}
and a similar expression for the individual ion Hamiltonians $\hat{H}_{i,k}^{(z)}$ under the substitution $\{M_p,\Omega_{zp}',\hat{P}_{zp},\delta{\hat{R}}_{zp}\}\to\{M_i,\Omega_{zi,k}',\hat{P}_{zi,k},\delta{\hat{R}}_{zi,k}\}$. The dissipative dynamics within this subspace are given by the terms within $\mathcal{D}_p(\hat\rho)$ and $\mathcal{D}_{i,k}(\hat\rho)$ containing operators corresponding to motion along $z$. 

\begin{figure}[t!]
    \centering
    \includegraphics[width=1\linewidth]{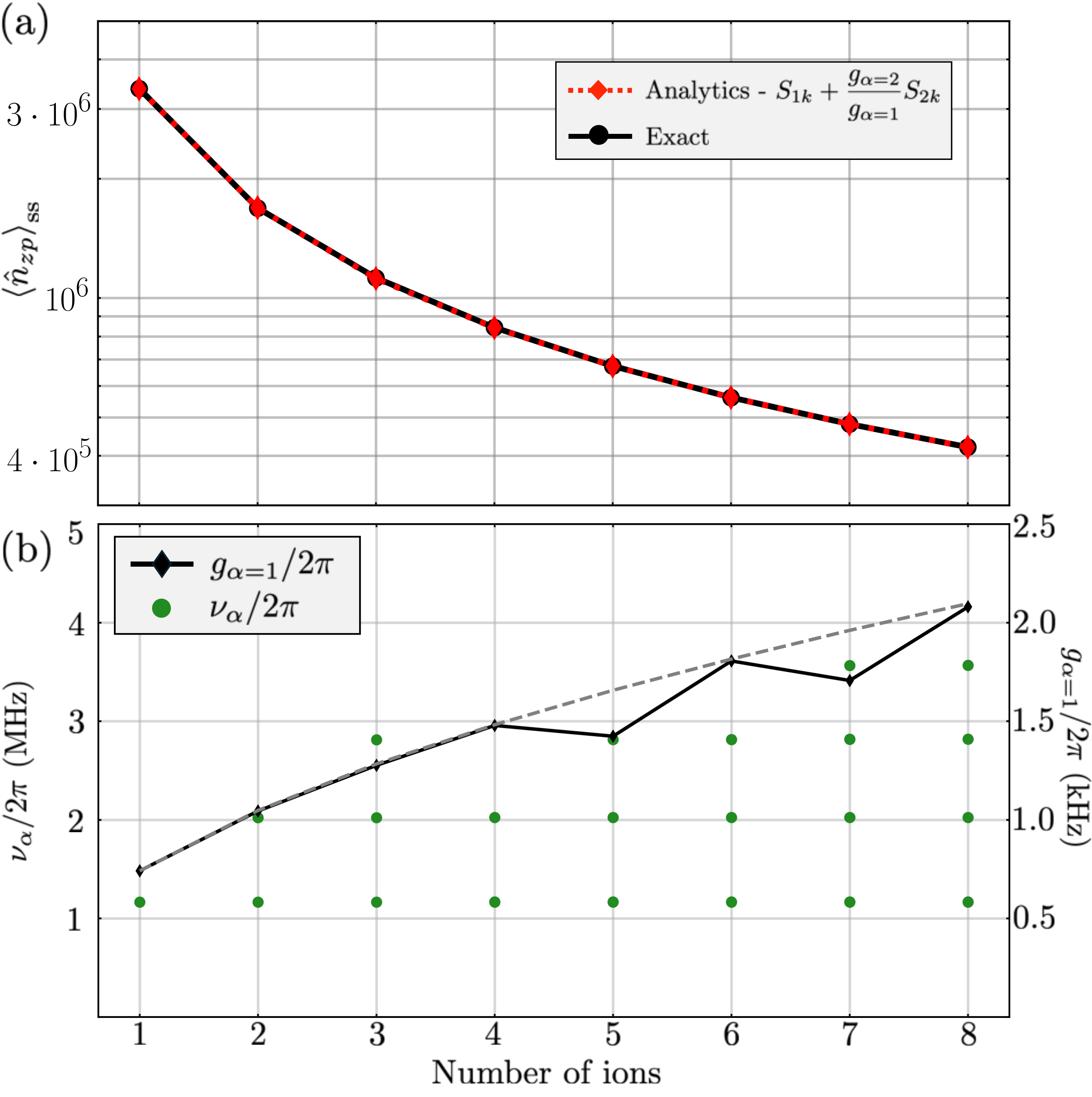}
    \caption{(a) Steady-state motional occupation of the nanoparticle along the $z$-axis, $\expval{\hat{n}_{zp}}_{\rm ss}$, versus ion number $N$, for the best-performing equilibrium configuration for each $N$. Parameters are taken from Table~\ref{tab:parameters}, with trap-displacement-like noise set to $\Gamma^{\rm td}_z=0$. The exact solution (black) is compared to a model where only the center-of-mass mode of the chain couples to the nanoparticle motion (red,  Eq.~\eqref{gammaDOP2}). (b) Normal mode frequencies of the $N-$ion system (green circles) and coupling rate between nanoparticle and normal mode $\alpha=1$ (black). The dashed gray curve shows a $\sqrt{N}$ scaling.}
    \label{fig:n,Omega,g-vs-N}
\end{figure}

Using Eq.~(\ref{HN_onlyz}) and the equilibrium positions obtained above, we numerically compute the steady-state occupation number of the nanoparticle motion along the $z$-axis, $\langle \hat n_{zp}\rangle_{\rm ss}$. 
We choose the best equilibrium configurations, namely those for which the steady-state occupation number $\langle \hat n_{zp}\rangle_{\rm ss}$ is lowest, and show the corresponding occupation number as a function of the number of ions in Fig.~\ref{fig:n,Omega,g-vs-N}(a) (black curve) in the absence of trap-displacement noise, $\Gamma^{\rm td}_z=0$. The nanoparticle occupation number decreases as $\sim1/N$, reaching an occupation $\langle \hat n_{zp}\rangle_{\rm ss}=4.2\times 10^5$ for $N=8$, corresponding to a motional temperature of $21~\mathrm{mK}$, a factor $\simeq 8$ lower than the single-ion case shown on the left-hand side in Fig.~\ref{fig:n-vs-gammap,Gammap}. 

To gain a deeper understanding of the sympathetic cooling process, we perform a symplectic diagonalisation of the ion Hamiltonian to obtain its normal modes. That is, we define a transformation
\begin{equation}\label{Xdiag}
    \hat{Q}_\alpha \equiv \sum_{k=1}^N \sqrt{M_i}  S_{\alpha k} \delta\hat{R}_{zi,k},
\end{equation}
and
\begin{equation}\label{Pdiag}
     \hat{\Pi}_\alpha \equiv \sum_{k=1}^N   \frac{S_{\alpha k} \hat{P}_{zi,k}}{\sqrt{M_i}},
\end{equation}
where the new coordinates obey canonical commutation relations, $\left[\hat{Q}_\alpha,\hat{\Pi}_{\alpha'}\right]=i\hbar\delta_{\alpha\alpha'}$, and where $S_{\alpha k}$ is an orthogonal matrix that diagonalises the ion part of Eq.~\eqref{HN_onlyz} and that is calculated numerically. In terms of these new coordinates, Eq.~\eqref{HN_onlyz} takes the form
\begin{multline}\label{H_tot_z}
    \hat{H}_{\rm tot}^{(z)} = \hat{H}_{p}^{(z)}+\sum_{\alpha=1}^N\Bigg(\frac{\hat{\Pi}_\alpha^2}{2} + \frac{1}{2}\nu_\alpha^2\hat{Q}_\alpha^2\Bigg)+\\
    + \delta{\hat{R}}_{zp}\sum_{\alpha=1}^N\hat Q_\alpha\sum_{k=1}^N \frac{S_{\alpha k} \big( \bar{N}_{kp}\big)_{33}}{\sqrt{M_i}}  .
\end{multline}
The coordinates $\hat{Q}_\alpha$ correspond to the normal modes of the $N-$ion ensemble, with the rows of $S_{\alpha k}$ describing the amplitude of motion of ion $k$ in mode $\alpha$.
Equation~\eqref{H_tot_z} thus describes a set of independent modes, each coupled to the nanoparticle motional coordinate at a rate
\begin{equation}
    g_\alpha\equiv\frac{1}{2}\sqrt{\frac{1}{M_i M_p\Omega_{zp}'\nu_\alpha}} \sum_{k=1}^N S_{ \alpha k} \big( \bar{N}_{kp}\big)_{33} .
\end{equation}
The normal mode frequencies $\nu_\alpha$ are shown in Fig.~\ref{fig:n,Omega,g-vs-N}(b). They have a very weak dependence on $N$, as has already been pointed out in the literature~\cite{James_1998}.

\begin{figure}[t!]
    \centering
    \includegraphics[width=1\linewidth]{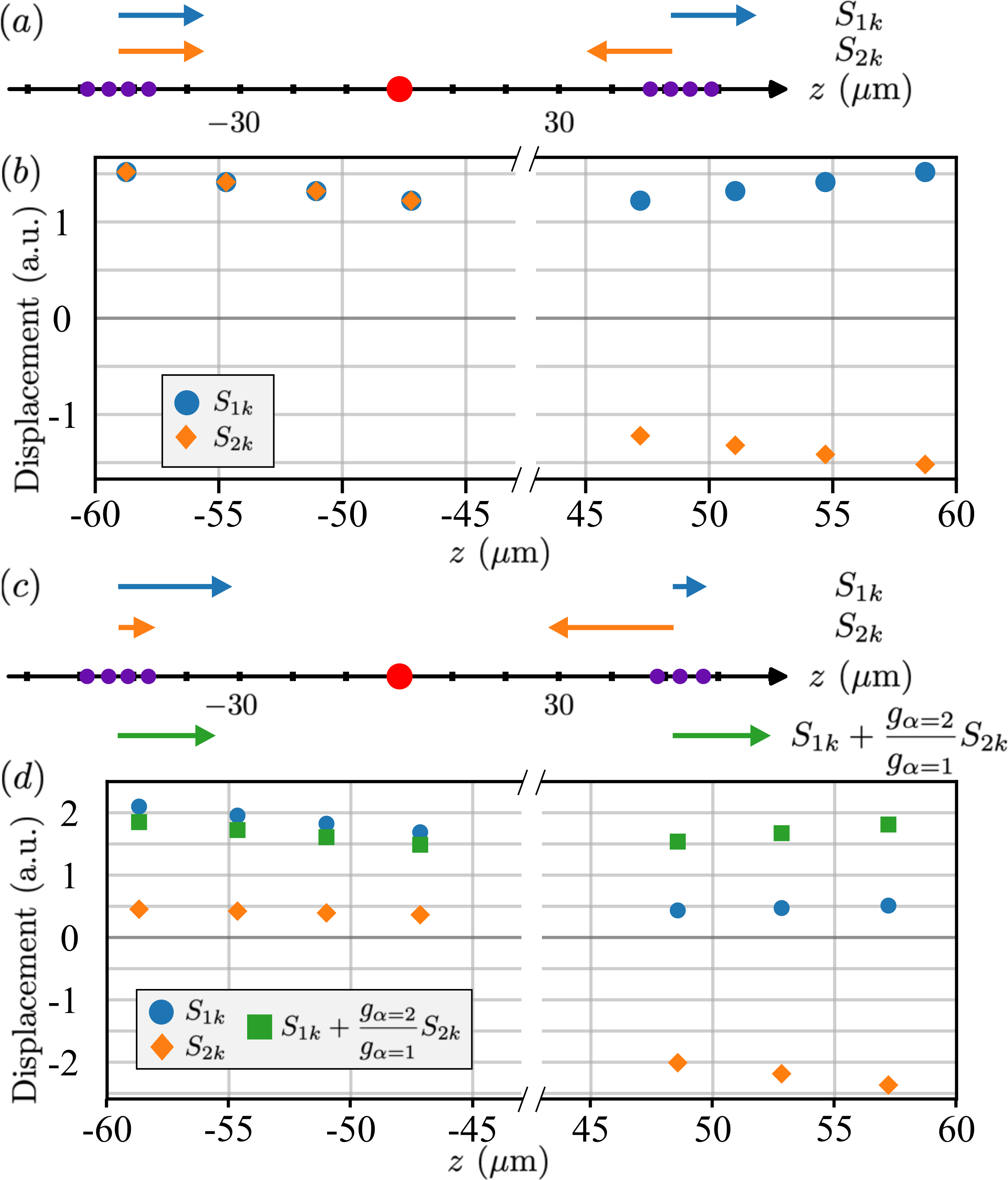}
    \caption{
Ion displacements for the first two normal modes, $S_{1k}$ (blue) and $S_{2k}$ (orange) for $N=8$ (panels a-b) and $N=7$ ions (panels c-d). In panels (a,c) red and purple disks represent nanoparticle and ions respectively, and the arrow lengths (not to scale) indicate the mean ion displacement on each chain. The nanoparticle motion couples predominantly to the center-of-mass mode of the ion ensemble, which for $N$ odd is a linear combination of the $\alpha=1$ and $\alpha=2$ modes (green arrows / dots).}
    \label{fig:normal-modes}
\end{figure}

The lowest collective mode frequency $\nu_\alpha \approx 2\pi\times 1.2~\mathrm{MHz}$ corresponds to a pair of quasi-degenerate modes which we label $\alpha=1$ and $\alpha=2$.
The mode displacements $S_{\alpha k}$ for these two modes are shown in Fig.~\ref{fig:normal-modes} for $N$ even (panels a-b) and $N$ odd (panels c-d). 
For $N$ even, the equilibrium configuration corresponds to the ions being arranged symmetrically in two chains on opposite sides of the nanoparticle, each containing $N/2$ ions (Fig.~\ref{fig:normal-modes}a). In this case, due to the inversion symmetry, the mode $\alpha=1$ corresponds to the center-of-mass mode where all ions oscillate in phase with approximately equal amplitudes, whereas the $\alpha=2$ mode corresponds to a breathing mode where the two ion chains oscillate with opposite phases. In contrast, for $N$ odd the equilibrium configuration is asymmetric with one ion chain having one ion less than the other (Fig.~\ref{fig:normal-modes}c). This asymmetry manifests in different motional amplitudes for each chain. 
In this case, the center-of-mass mode is a linear combination of both modes $\alpha=1$ and $\alpha=2$ with  weights $g_{\alpha=1}$ and $g_{\alpha=2}$. We expect the sympathetic cooling to be dominated by the center-of-mass mode as the forces exerted by each ion onto the nanoparticle add up constructively at all times. 
This is confirmed by the red curve in Fig.~\ref{fig:n,Omega,g-vs-N}(a), which shows the approximate occupation, $\langle \hat n_{zp}\rangle_{\rm ss} \approx \Gamma_z^{\rm gas}/\gamma_{p}^{\rm eff,N}$, using an effective cooling rate similar to Eq.~(\ref{gammaDOP}) but including only normal modes that contribute to the center-of-mass mode,
\begin{multline}\label{gammaDOP2}
    \gamma_{p}^{\rm eff,N} =\gamma_{p,\alpha=1}^{\rm eff}+ \gamma_{p,\alpha=2}^{\rm eff}\delta_{N,\rm odd} \\ \approx  \gamma^{\rm dop}  \frac{4 \Omega_{zp}^{'}}{ \nu_1^3 }  \times\left\lbrace\begin{array}{cc}
        g_{\alpha=1}^2 & \hspace{0.05cm} \text{($N$ even)} \\
        g_{\alpha=1}^2+g_{\alpha=2}^2 & \hspace{0.05cm} \text{($N$ odd)}
    \end{array}\right.
\end{multline}
In the last step we have used $\nu_{\alpha=1} \approx \nu_{\alpha=2}$ and approximated the cooling rates of both normal modes as the single-ion cooling rate $\gamma^{\rm dop}$, an approximation that is valid for any mode $\alpha$ fulfilling $\vert\nu_{\alpha}-\Omega_{zi}'\vert \ll \sqrt{\nu_{\alpha}\Omega_{zi}'}$.

For small $N$, one can find stable configurations where all ions are located on the same side of the nanoparticle. In this case, a similar argument can be made, the only difference being that the lowest normal-mode frequency $\nu_\alpha \approx 2\pi\times 1.2~\mathrm{MHz}$ corresponds to the non-degenerate center-of-mass mode of the ion chain for both $N$ even and $N$ odd. As a consequence the coupling between nanoparticle and other normal modes is also small for both $N$ even and $N$ odd, specifically $g_{\alpha=2}^2 \ll g_{\alpha = 1}^2$. Because of this, the expression for the effective cooling rate, Eq.~\eqref{gammaDOP2}, remains valid.

Equation~\eqref{gammaDOP2} provides a useful resource to estimate the cooling rate expected in general setups, as well as its scaling with $N$.
Specifically, the only factors in Eq.~\eqref{gammaDOP2} that significantly depend on the ion number $N$ are the normal mode coupling rates $g_\alpha$, which display the $\sqrt{N}$ scaling characteristic of homogeneous coupling to multiple degrees of freedom (see, e.g., the coupling $g_{\alpha=1}$ displayed in Fig.~\ref{fig:n,Omega,g-vs-N}(b)). This results in a scaling $\gamma_{p}^{\rm eff,N}\sim N$ for the cooling rate and fully explains the scaling $\langle \hat n_{zp}\rangle_{\rm ss}\sim 1/N$ for the occupation observed in Fig.~\ref{fig:n,Omega,g-vs-N}(a).    This scaling suggests that sympathetic cooling is unlikely to enable cooling of the nanoparticle motion to the ground state, as co-trapping $\sim 10^6-10^7$ ions is technically challenging. Our results thus highlight the relevance of devising new methods to increase sympathetic cooling efficiency.

\section{Conclusion}\label{sec:conclusion}

We have derived a theory to describe the quantum dynamics of a levitated nanoparticle co-trapped with an ion ensemble in a dual-frequency linear Paul trap. 
We have used it to make quantitative predictions about ion-based sympathetic cooling in levitodynamics and to provide analytical estimates of the nanoparticle occupations, which apply to a wide variety of electric trap geometries and parameters. 
Our theory predicts that, if external noise is suppressed in state-of-the-art experiments, the nanoparticle motion can be sympathetically cooled 
 below $1~\mathrm{K}$ using a single ion, and below $25~\mathrm{mK}$ using the center-of-mass mode of an ensemble of $N=8$ ions. We have identified the main limitations of this scheme to be high trap-displacement noise rates $\Gamma_j^{\rm td}$ and, most importantly, the vast difference between ion and nanoparticle mechanical frequencies, $\Omega_{ji}'$ and $\Omega_{jp}'$. To obtain insight into these limitations, we 
can set $\Gamma_j^{\rm td}=0$ and equal mechanical frequencies, $\Omega_{ji}'=\Omega_{jp}'=2\pi\times 1~\mathrm{kHz}$ in our model, while keeping the same values used throughout this work for the remaining parameters~\footnote{Note that this is only a thought experiment, since having equal mechanical frequencies would not correspond to a stable dual-frequency trap for both particles. Also, changing the frequencies would necessarily change other parameters as well such as the coupling rate $g$, which we leave unchanged for this estimate.}. This choice results in a predicted steady-state nanoparticle occupation of $119$ phonons using a single ion, in the absence of optical feedback. Future efforts toward ground-state cooling should thus be aimed not only at reducing unwanted noise, but also at bridging this frequency gap, for instance through coupling both particle and ions to the same optical cavity~\cite{Pflanzer_2013}. A second option is to modulate the distance between ion and nanoparticle, and thus their Coulomb coupling, at a frequency $\Omega_{ji}'-\Omega_{jp}'$ in order to parametrically amplify the sympathetic cooling. This modulation could be achieved in free space, i.e., without the need for cavities, by, e.g., applying an AC voltage to one of the endcap electrodes
{\color{black}
~\footnote{{\color{black}Since $|g|\ll\Omega_{ji}'-\Omega_{jp}'$, the modes hybridize very weakly, e.g., the bare nanoparticle mode has a fraction of $\sim 10^{-6}$ in the ion-like normal mode, so it is possible to address predominantly ion motion with a single drive tone.}}. Lastly, although rotations do not play a relevant role in our system (see Appendix~\ref{Appendixmultipoles}), it would be interesting to explore if librations with megahertz-range frequencies of spinning anisotropic particles, as demonstrated in~\cite{Zielinska_2024}, could serve to bridge the frequency gap. Finally, we note that nonclassical behaviour can in principle be observed even without ground-state cooling; for example, a recent proposal to prepare non-Gaussian quantum states requires only cooling to $\langle\hat{n}_{jp}\rangle\sim 10^3$~\cite{Schut_2025}.
}

Beyond sympathetic cooling, the theory we have developed has enabled us to provide analytical expressions for all system parameters, including the secular motional frequencies, coupling rates, and equilibrium positions of ion and nanoparticle. 
Our work thus provides the theoretical toolbox needed to describe ion-nanoparticle levitodynamics experiments in the quantum regime{\color{black}, going past the semiclassical regime of current experiments}. 
Moreover, it opens the door to  assess the feasibility of protocols to prepare quantum motional states of electrically levitated nanoparticles without the use of light.

\begin{acknowledgments}
We are especially grateful to Andreas Deutschmann-Olek for many insightful and stimulating discussions. This research was funded in whole or in part by the Austrian Science Fund (FWF) [10.55776/COE1]. For open access purposes, the author has applied a CC BY public copyright license to any author accepted manuscript version arising from this submission.

\end{acknowledgments}

S. G. and C. G. B. contributed equally to this work.
\newpage

\appendix

\section{Exact expression for nanoparticle occupation}\label{Appendixformula}

The nanoparticle's phonon occupation number obtained from Eq.~\eqref{covmatstatic} can be cast as
\begin{align}
\begin{aligned}
    \expval{\hat{n}_{jp}}_{\rm ss} &= 
    \frac{
    32\tilde{g}_j^4 n_1 - k \chi \tilde{\gamma}^{\rm dop} \tilde{\Gamma}_{jp} n_2 + 4 \tilde{g}_j^2 \tilde{\gamma}^{\rm dop} n_3
    }{
    \tilde{\gamma}^{\rm dop} \left( 4 k \chi - 64 \tilde{\gamma}^{\rm dop 2} \tilde{g}_j^2 \tilde{\Omega}_{jp} \right) \left( 64 \gamma_+^2 \tilde{g}_j^2 \tilde{\Omega}_{jp} + \gamma_{\times} n_2 \right)
    },
\end{aligned}
\end{align}
where where variables with tildes have been normalized to the ion's mechanical frequency, e.g. $\tilde\gamma^{\rm dop}\equiv \gamma^{\rm dop}/\Omega_{ji}'$, and with
\begin{widetext}
\begin{multline}
    n_1 \equiv \left(\gamma_{\times} \gamma_+ h_3 + \tilde{\gamma}^{\rm dop 3}\left(4+\gamma_+\tilde{\gamma}^{\rm dop 2} \right)\right) \left( \gamma_{\times} + 2 \tilde{\gamma}^{\rm dop} \tilde{\Gamma}_{jp} \right)
    - \tilde{A} \left(\gamma_{\times}^2 h_2 + 8 k \tilde{\gamma}^{\rm dop 2} h_{1/2} \right) \tilde{\Omega}_{jp} \\
    + 4 \tilde{\gamma}^{\rm dop} \left( h_2^2 - 6 \tilde{\gamma}^{\rm dop} \tilde{\Gamma}_{jp} h_{4/3} \right) \tilde{\Omega}_{jp}^2
    - 4 \tilde{A} \tilde{\gamma}^{\rm dop 2} h_2 \tilde{\Omega}_{jp}^3,
\end{multline}
\begin{align}
    n_2 \equiv  \left( \gamma_+^2 + 4 \left( 1 + \tilde{\Omega}_{jp}^2 \right) \right)^2 - 64 \tilde{\Omega}_{jp}^2,
\end{align}
and
\begin{align}
\begin{split}
    n_3 \equiv 
    & k \gamma_{\times}^2 \gamma_+ \left(4 + \gamma_+^2 \right) \tilde{A} - 8 k \gamma_{\times}^2 \gamma_+ \tilde{\gamma}^{\rm dop} \tilde{\Omega}_{jp}
    -4 \left( -4 \gamma_{\times} \gamma_+ h_{-2} + \gamma_{\times}^2 \gamma_+ h_2 + 3 \gamma_+ \tilde{\gamma}^4 h_{1/3} + 16 \tilde{\gamma}^{\rm dop 3} + 8 \gamma_+ \tilde{\gamma}^{\rm dop 4} \right) \tilde{\Gamma}_{jp} \tilde{\Omega}_{jp} \\
    &+ 8 k \gamma_+ \tilde{A} \left( 2 k \tilde{\gamma}^{\rm dop 2} + \gamma_{\times} h_1 \right) \tilde{\Omega}_{jp}^2
    - 8 \tilde{\gamma}^{\rm dop} \left( 4 k \gamma_+ \tilde{\gamma}^{\rm dop 2} + 2 \left( -4 h_3 + 2 \gamma_{\times} h_{3/2} + \tilde{\gamma}^{\rm dop 4} \right) \tilde{\Gamma}_{jp} \right) \tilde{\Omega}_{jp}^3 \\
    &+ 16 k \gamma_+ \tilde{A} \tilde{\gamma}^{\rm dop 2} \tilde{\Omega}_{jp}^4 - 64 \tilde{\gamma}^{\rm dop 3} \tilde{\Gamma}_{jp} \tilde{\Omega}_{jp}^5,
\end{split}
\end{align}
\end{widetext}
and where we have defined $\tilde{A} \equiv \tilde{\gamma}^{\rm dop} + 2 \tilde{\Gamma}_{j}^{\rm dop}$, $k \equiv (\tilde\gamma^{\rm dop}/2)^2 + 1$, $\gamma_+ = \tilde{\gamma}_{p} + \tilde{\gamma}^{\rm dop}$, $\gamma_\times = \tilde{\gamma}_{p} \tilde{\gamma}^{\rm dop}$, $h_\eta \equiv \gamma_{\times} + \eta \gamma_+^2$, and $\chi \equiv \gamma_{\times}^2 + 4 \tilde{\gamma}^{\rm dop 2} \tilde{\Omega}_{jp}^2$.
{\color{black}
\section{Rotation-translation couplings due to charge distribution multipoles}\label{Appendixmultipoles}

The nanoparticle has a finite number of discrete charges, so its centers of mass and charge do not coincide exactly. We assume that point-like elementary charges are randomly distributed on the nanoparticle surface, which allows us to calculate the average residual dipole and quadrupole moments with respect to the center-of-mass $\mathbf{D}_p$,
\begin{equation}
    \mathbf{p}=e\sum_j\mathbf{r}_j, \quad \mathbb{Q}=e\sum_j(3\mathbf{r}_j\otimes\mathbf{r}_j-r_j^2\mathbf{1}), \quad \mathbf{r}_j=\mathbf{R}_j-\mathbf{D}_p.
\end{equation}
For the parameters of this work,
\begin{equation}
    \langle|\mathbf{p}|\rangle\sim eR_p\sqrt{Q_p/e}\sim 6\times10^{-25}\,\mathrm {C\cdot m}
\end{equation}
\begin{equation}
    \langle|Q_j|\rangle\sim eR_p^2\sqrt{Q_p/e}\sim 8\times 10^{-32}\,\rm C\cdot m^2,
    \label{estimate_pQ}
\end{equation}
where $Q_j$ denotes the eigenvalues of $\mathbb{Q}$. The potential energy in an external potential $\phi(\mathbf{R}, t)$ is given by the multipole expansion
\begin{multline}
    U(\mathbf{D}_p, \mathbf{\Omega}, t)=Q_p\phi(\mathbf{D}_p, t)+\mathbf{p}(\mathbf{\Omega})\cdot\nabla\phi(\mathbf{D}_p, t)+\\+\frac{1}{6}\Tr\Big[\mathbb{Q}(\mathbf{\Omega})\mathbb{H}(\phi(\mathbf{D}_p, t))\Big],
\end{multline}
where $\mathbf{\Omega}$ denotes the angular coordinates and $\mathbb{H(\cdot)}$ denotes the Hessian. At the equilibrium position the field vanishes, $\nabla\phi(\mathbf{D}_{p,0})=0$, and we can calculate the difference
\begin{equation}
\begin{split}
    \Delta U(\mathbf{D}_{p,0})&=\max_{\mathbf{\Omega},t}U(\mathbf{D}_{p,0}, \mathbf{\Omega},t)-\min_{\mathbf{\Omega},t}U(\mathbf{D}_{p,0}, \mathbf{\Omega},t)\\
    &=\frac{1}{6}\Big(\max_{\mathbf{\Omega},t}\Tr\Big[\mathbb{Q}(\mathbf{\Omega})\mathbb{H}(\phi(\mathbf{D}_{p,0}, t))\Big]-\\
    &-\min_{\mathbf{\Omega},t}\Tr\Big[\mathbb{Q}(\mathbf{\Omega})\mathbb{H}(\phi(\mathbf{D}_{p,0}, t))\Big]]\Big),
\end{split}
\end{equation}
that corresponds to the size of the potential barrier for the particle orientation. When we plug in the parameter values from Table~\ref{tab:parameters} and estimates~\eqref{estimate_pQ}, for the relevant trapping tone at $7$~kHz we get $\Delta U(\mathbf{D}_{p,0})\sim 7\times10^{-24}\,\mathrm{J}\sim k_B\cdot 0.5\,\rm K$. Thus, at room temperature the nanoparticle will not be orientationally trapped. Reference~\cite{Perdriat2024} studied rotational motion in a conceptually similar setup and observed locking of rotational motion to the frequency of the trapping voltage, although with a different particle geometry. Here, since the typical thermal angular velocity is $\omega_{\rm th}\sim\sqrt{k_BT/I}\sim 2\pi\times30\,\rm kHz$ (where $I=(2/5)\cdot M_pR_p^2$), the nanoparticle will not synchronize with the 7~kHz trapping tone, but will behave like a weakly perturbed free rotor instead. The angular velocity is large enough that we can adiabatically eliminate rotational degrees of freedom, so their influence on translational dynamics is negligible. The damping rate $\gamma^{\rm gas}$ is very small~\cite{vanderLaan_2021, vanderLaan_2022, Martinetz_2018, Stickler_2021}, so on the experimentally relevant timescales the direction of the angular velocity vector is fixed, not thermally randomized.
}

\bibliographystyle{apsrev4-2} 
\bibliography{bibliography} 

@article{Pogorelov_2021,
  title = {Compact Ion-Trap Quantum Computing Demonstrator},
  author = {Pogorelov, I. and Feldker, T. and Marciniak, Ch. D. and Postler, L. and Jacob, G. and Krieglsteiner, O. and Podlesnic, V. and Meth, M. and Negnevitsky, V. and Stadler, M. and H\"ofer, B. and W\"achter, C. and Lakhmanskiy, K. and Blatt, R. and Schindler, P. and Monz, T.},
  journal = {PRX Quantum},
  volume = {2},
  issue = {2},
  pages = {020343},
  numpages = {23},
  year = {2021},
  month = {Jun},
  publisher = {American Physical Society},
  doi = {10.1103/PRXQuantum.2.020343},
  url = {https://link.aps.org/doi/10.1103/PRXQuantum.2.020343}
}

@article{Enzer_2000,
  title = {Observation of Power-Law Scaling for Phase Transitions in Linear Trapped Ion Crystals},
  author = {Enzer, D. G. and Schauer, M. M. and Gomez, J. J. and Gulley, M. S. and Holzscheiter, M. H. and Kwiat, P. G. and Lamoreaux, S. K. and Peterson, C. G. and Sandberg, V. D. and Tupa, D. and White, A. G. and Hughes, R. J. and James, D. F. V.},
  journal = {Phys. Rev. Lett.},
  volume = {85},
  issue = {12},
  pages = {2466--2469},
  numpages = {0},
  year = {2000},
  month = {Sep},
  publisher = {American Physical Society},
  doi = {10.1103/PhysRevLett.85.2466},
  url = {https://link.aps.org/doi/10.1103/PhysRevLett.85.2466}
}

@article{Teller_2023,
    author = {Teller, Markus and Messerer, Viktor and Schüppert, Klemens and Zou, Yueyang and Fioretto, Dario A. and Galli, Maria and Holz, Philip C. and Reichel, Jakob and Northup, Tracy E.},
    title = {Integrating a fiber cavity into a wheel trap for strong ion–cavity coupling},
    journal = {AVS Quantum Science},
    volume = {5},
    number = {1},
    pages = {012001},
    year = {2023},
    month = {01},
    issn = {2639-0213},
    doi = {10.1116/5.0121534},
    url = {https://doi.org/10.1116/5.0121534},
}

@article{Millen_2020,
	author = {Millen, James and Monteiro, Tania S and Pettit, Robert and Vamivakas, A Nick},
	doi = {10.1088/1361-6633/ab6100},
	journal = {Reports on Progress in Physics},
	month = {jan},
	number = {2},
	pages = {026401},
	publisher = {IOP Publishing},
	title = {Optomechanics with levitated particles},
	url = {https://dx.doi.org/10.1088/1361-6633/ab6100},
	volume = {83},
	year = {2020}}

@article{CGB_2021,
author = {C. Gonzalez-Ballestero  and M. Aspelmeyer  and L. Novotny  and R. Quidant  and O. Romero-Isart },
title = {Levitodynamics: Levitation and control of microscopic objects in vacuum},
journal = {Science},
volume = {374},
number = {6564},
pages = {eabg3027},
year = {2021},
doi = {10.1126/science.abg3027},
URL = {https://www.science.org/doi/abs/10.1126/science.abg3027}}

@article{Delic_2020,
author = {Uroš Delić  and Manuel Reisenbauer  and Kahan Dare  and David Grass  and Vladan Vuletić  and Nikolai Kiesel  and Markus Aspelmeyer },
title = {Cooling of a levitated nanoparticle to the motional quantum ground state},
journal = {Science},
volume = {367},
number = {6480},
pages = {892-895},
year = {2020},
doi = {10.1126/science.aba3993},
URL = {https://www.science.org/doi/abs/10.1126/science.aba3993}}

@article{Tebbenjohanns_2021,
	author = {Tebbenjohanns, Felix and Mattana, M. Luisa and Rossi, Massimiliano and Frimmer, Martin and Novotny, Lukas},
	date = {2021/07/01},
	id = {Tebbenjohanns2021},
	isbn = {1476-4687},
	journal = {Nature},
	number = {7867},
	pages = {378--382},
	title = {Quantum control of a nanoparticle optically levitated in cryogenic free space},
	url = {https://doi.org/10.1038/s41586-021-03617-w},
	volume = {595},
	year = {2021}}

@article{Magrini_2021,
	author = {Magrini, Lorenzo and Rosenzweig, Philipp and Bach, Constanze and Deutschmann-Olek, Andreas and Hofer, Sebastian G. and Hong, Sungkun and Kiesel, Nikolai and Kugi, Andreas and Aspelmeyer, Markus},
	date = {2021/07/01},
	id = {Magrini2021},
	isbn = {1476-4687},
	journal = {Nature},
	number = {7867},
	pages = {373--377},
	title = {Real-time optimal quantum control of mechanical motion at room temperature},
	url = {https://doi.org/10.1038/s41586-021-03602-3},
	volume = {595},
	year = {2021}}

@article{Kamba_2022,
	author = {Mitsuyoshi Kamba and Ryoga Shimizu and Kiyotaka Aikawa},
	doi = {10.1364/OE.462921},
	journal = {Opt. Express},
	month = {Jul},
	number = {15},
	pages = {26716--26727},
	publisher = {Optica Publishing Group},
	title = {Optical cold damping of neutral nanoparticles near the ground state in an optical lattice},
	url = {https://opg.optica.org/oe/abstract.cfm?URI=oe-30-15-26716},
	volume = {30},
	year = {2022}}

@article{Ranfagni_2022,
  title = {Two-dimensional quantum motion of a levitated nanosphere},
  author = {Ranfagni, A. and B\o{}rkje, K. and Marino, F. and Marin, F.},
  journal = {Phys. Rev. Res.},
  volume = {4},
  issue = {3},
  pages = {033051},
  numpages = {10},
  year = {2022},
  month = {Jul},
  publisher = {American Physical Society},
  doi = {10.1103/PhysRevResearch.4.033051},
  url = {https://link.aps.org/doi/10.1103/PhysRevResearch.4.033051}
}

@article{Piotrowski_2023,
	author = {Piotrowski, Johannes and Windey, Dominik and Vijayan, Jayadev and Gonzalez-Ballestero, Carlos and de los R{\'\i}os Sommer, Andr{\'e}s and Meyer, Nadine and Quidant, Romain and Romero-Isart, Oriol and Reimann, Ren{\'e} and Novotny, Lukas},
	date = {2023/07/01},
	id = {Piotrowski2023},
	isbn = {1745-2481},
	journal = {Nature Physics},
	number = {7},
	pages = {1009--1013},
	title = {Simultaneous ground-state cooling of two mechanical modes of a levitated nanoparticle},
	url = {https://doi.org/10.1038/s41567-023-01956-1},
	volume = {19},
	year = {2023}}

@article{Dania_2025,
	author = {Dania, Lorenzo and Kremer, Oscar Schmitt and Piotrowski, Johannes and Candoli, Davide and Vijayan, Jayadev and Romero-Isart, Oriol and Gonzalez-Ballestero, Carlos and Novotny, Lukas and Frimmer, Martin},
	date = {2025/08/06},
	id = {Dania2025},
	isbn = {1745-2481},
	journal = {Nature Physics},
	title = {High-purity quantum optomechanics at room temperature},
	url = {https://doi.org/10.1038/s41567-025-02976-9},
	year = {2025}}

@article{Rossi_2024,
  title = {Quantum Delocalization of a Levitated Nanoparticle},
  author = {Rossi, M. and Militaru, A. and Carlon Zambon, N. and Riera-Campeny, A. and Romero-Isart, O. and Frimmer, M. and Novotny, L.},
  journal = {Phys. Rev. Lett.},
  volume = {135},
  issue = {8},
  pages = {083601},
  numpages = {7},
  year = {2025},
  month = {Aug},
  publisher = {American Physical Society},
  doi = {10.1103/2yzc-fsm3},
  url = {https://link.aps.org/doi/10.1103/2yzc-fsm3}
}

@article{
Kamba_2025,
author = {Mitsuyoshi Kamba  and Naoki Hara  and Kiyotaka Aikawa },
title = {Quantum squeezing of a levitated nanomechanical oscillator},
journal = {Science},
volume = {389},
number = {6766},
pages = {1225-1228},
year = {2025},
doi = {10.1126/science.ady4652},
URL = {https://www.science.org/doi/abs/10.1126/science.ady4652}}

@article{Jain_2016,
  title = {Direct Measurement of Photon Recoil from a Levitated Nanoparticle},
  author = {Jain, Vijay and Gieseler, Jan and Moritz, Clemens and Dellago, Christoph and Quidant, Romain and Novotny, Lukas},
  journal = {Phys. Rev. Lett.},
  volume = {116},
  issue = {24},
  pages = {243601},
  numpages = {5},
  year = {2016},
  month = {Jun},
  publisher = {American Physical Society},
  doi = {10.1103/PhysRevLett.116.243601},
  url = {https://link.aps.org/doi/10.1103/PhysRevLett.116.243601}
}

@article{ORI_2011_july,
  title = {Large Quantum Superpositions and Interference of Massive Nanometer-Sized Objects},
  author = {Romero-Isart, O. and Pflanzer, A. C. and Blaser, F. and Kaltenbaek, R. and Kiesel, N. and Aspelmeyer, M. and Cirac, J. I.},
  journal = {Phys. Rev. Lett.},
  volume = {107},
  issue = {2},
  pages = {020405},
  numpages = {4},
  year = {2011},
  month = {Jul},
  publisher = {American Physical Society},
  doi = {10.1103/PhysRevLett.107.020405},
  url = {https://link.aps.org/doi/10.1103/PhysRevLett.107.020405}
}

@article{ORI_2011_nov,
  title = {Quantum superposition of massive objects and collapse models},
  author = {Romero-Isart, Oriol},
  journal = {Phys. Rev. A},
  volume = {84},
  issue = {5},
  pages = {052121},
  numpages = {17},
  year = {2011},
  month = {Nov},
  publisher = {American Physical Society},
  doi = {10.1103/PhysRevA.84.052121},
  url = {https://link.aps.org/doi/10.1103/PhysRevA.84.052121}
}

@article{Bateman_2014,
	author = {Bateman, James and Nimmrichter, Stefan and Hornberger, Klaus and Ulbricht, Hendrik},
	date = {2014/09/02},
	doi = {10.1038/ncomms5788},
	id = {Bateman2014},
	isbn = {2041-1723},
	journal = {Nature Communications},
	number = {1},
	pages = {4788},
	title = {Near-field interferometry of a free-falling nanoparticle from a point-like source},
	url = {https://doi.org/10.1038/ncomms5788},
	volume = {5},
	year = {2014}}

@article{Neumeier_2024,
	author = {Lukas Neumeier and Mario A. Ciampini and Oriol Romero-Isart and Markus Aspelmeyer and Nikolai Kiesel},
	doi = {10.1073/pnas.2306953121},
	journal = {Proceedings of the National Academy of Sciences},
	number = {4},
	pages = {e2306953121},
	title = {Fast quantum interference of a nanoparticle via optical potential control},
	url = {https://www.pnas.org/doi/abs/10.1073/pnas.2306953121},
	volume = {121},
	year = {2024}}

@article{MRL_2024,
  title = {Macroscopic Quantum Superpositions via Dynamics in a Wide Double-Well Potential},
  author = {Roda-Llordes, M. and Riera-Campeny, A. and Candoli, D. and Grochowski, P. T. and Romero-Isart, O.},
  journal = {Phys. Rev. Lett.},
  volume = {132},
  issue = {2},
  pages = {023601},
  numpages = {7},
  year = {2024},
  month = {Jan},
  publisher = {American Physical Society},
  doi = {10.1103/PhysRevLett.132.023601},
  url = {https://link.aps.org/doi/10.1103/PhysRevLett.132.023601}
}

@article{James_1998,
	Author = {James, D. F. V. },
	Da = {1998/02/01},
	Doi = {10.1007/s003400050373},
	Id = {James1998},
	Isbn = {1432-0649},
	Journal = {Applied Physics B},
	Number = {2},
	Pages = {181--190},
	Title = {Quantum dynamics of cold trapped ions with application to quantum computation},
	Ty = {JOUR},
	Url = {https://doi.org/10.1007/s003400050373},
	Volume = {66},
	Year = {1998}}

@article{Casulleras_2024,
  title = {Optimization of static potentials for large delocalization and non-Gaussian quantum dynamics of levitated nanoparticles under decoherence},
  author = {Casulleras, Silvia and Grochowski, Piotr T. and Romero-Isart, Oriol},
  journal = {Phys. Rev. A},
  volume = {110},
  issue = {3},
  pages = {033511},
  numpages = {10},
  year = {2024},
  month = {Sep},
  publisher = {American Physical Society},
  doi = {10.1103/PhysRevA.110.033511},
  url = {https://link.aps.org/doi/10.1103/PhysRevA.110.033511}
}

@article{ORI_2012,
  title = {Quantum Magnetomechanics with Levitating Superconducting Microspheres},
  author = {Romero-Isart, O. and Clemente, L. and Navau, C. and Sanchez, A. and Cirac, J. I.},
  journal = {Phys. Rev. Lett.},
  volume = {109},
  issue = {14},
  pages = {147205},
  numpages = {5},
  year = {2012},
  month = {Oct},
  publisher = {American Physical Society},
  doi = {10.1103/PhysRevLett.109.147205},
  url = {https://link.aps.org/doi/10.1103/PhysRevLett.109.147205}
}

@article{Millen_2015,
  title = {Cavity Cooling a Single Charged Levitated Nanosphere},
  author = {Millen, J. and Fonseca, P. Z. G. and Mavrogordatos, T. and Monteiro, T. S. and Barker, P. F.},
  journal = {Phys. Rev. Lett.},
  volume = {114},
  issue = {12},
  pages = {123602},
  numpages = {5},
  year = {2015},
  month = {Mar},
  publisher = {American Physical Society},
  doi = {10.1103/PhysRevLett.114.123602},
  url = {https://link.aps.org/doi/10.1103/PhysRevLett.114.123602}
}

@article{Fonseca_2016,
  title = {Nonlinear Dynamics and Strong Cavity Cooling of Levitated Nanoparticles},
  author = {Fonseca, P. Z. G. and Aranas, E. B. and Millen, J. and Monteiro, T. S. and Barker, P. F.},
  journal = {Phys. Rev. Lett.},
  volume = {117},
  issue = {17},
  pages = {173602},
  numpages = {5},
  year = {2016},
  month = {Oct},
  publisher = {American Physical Society},
  doi = {10.1103/PhysRevLett.117.173602},
  url = {https://link.aps.org/doi/10.1103/PhysRevLett.117.173602}
}

@article{Slezak_2018,
	author = {Slezak, Bradley R and Lewandowski, Charles W and Hsu, Jen-Feng and D'Urso, Brian},
	doi = {10.1088/1367-2630/aacac1},
	journal = {New Journal of Physics},
	month = {jun},
	number = {6},
	pages = {063028},
	publisher = {IOP Publishing},
	title = {Cooling the motion of a silica microsphere in a magneto-gravitational trap in ultra-high vacuum},
	url = {https://dx.doi.org/10.1088/1367-2630/aacac1},
	volume = {20},
	year = {2018}}

@article{Conangla_2020,
	author = {Conangla, Gerard P. and Rica, Ra{\'u}l A. and Quidant, Romain},
	date = {2020/08/12},
	doi = {10.1021/acs.nanolett.0c02025},
	isbn = {1530-6984},
	journal = {Nano Letters},
	journal1 = {Nano Letters},
	journal2 = {Nano Lett.},
	month = {08},
	n2 = {The levitation of condensed matter in vacuum allows the study of its physical properties under extreme isolation from the environment. It also offers a venue to investigate quantum mechanics with large systems, at the transition between the quantum and classical worlds. In this work, we study a novel hybrid levitation platform that combines a Paul trap with a weak but highly focused laser beam, a configuration that integrates a deep potential with excellent confinement and motion detection. We combine simulations and experiments to demonstrate the potential of this approach to extend vacuum trapping and interrogation to a broader range of nanomaterials, such as absorbing particles. We study the stability and dynamics of different specimens, such as fluorescent dielectric crystals and gold nanorods, and demonstrate stable trapping down to pressures of 1 mbar.},
	number = {8},
	pages = {6018--6023},
	publisher = {American Chemical Society},
	title = {Extending Vacuum Trapping to Absorbing Objects with Hybrid Paul-Optical Traps},
	type = {doi: 10.1021/acs.nanolett.0c02025},
	url = {https://doi.org/10.1021/acs.nanolett.0c02025},
	volume = {20},
	year = {2020},
	year1 = {2020}}

@article{Bykov_2022,
  title={Hybrid electro-optical trap for experiments with levitated particles in vacuum},
  author={Bykov, Dmitry S and Meusburger, Maximilian and Dania, Lorenzo and Northup, Tracy E},
url={https://pubs.aip.org/aip/rsi/article/93/7/073201/2848775/Hybrid-electro-optical-trap-for-experiments-with},
  journal={Review of scientific instruments},
  volume={93},
  number={7},
  year={2022},
  publisher={AIP Publishing}
}

@article{MGL_2023,
  title = {Superconducting Microsphere Magnetically Levitated in an Anharmonic Potential with Integrated Magnetic Readout},
  author = {Gutierrez Latorre, Mart\'{\i} and Higgins, Gerard and Paradkar, Achintya and Bauch, Thilo and Wieczorek, Witlef},
  journal = {Phys. Rev. Appl.},
  volume = {19},
  issue = {5},
  pages = {054047},
  numpages = {14},
  year = {2023},
  month = {May},
  publisher = {American Physical Society},
  doi = {10.1103/PhysRevApplied.19.054047},
  url = {https://link.aps.org/doi/10.1103/PhysRevApplied.19.054047}
}

@article{Hofer_2023,
  title = {High-$Q$ Magnetic Levitation and Control of Superconducting Microspheres at Millikelvin Temperatures},
  author = {Hofer, J. and Gross, R. and Higgins, G. and Huebl, H. and Kieler, O. F. and Kleiner, R. and Koelle, D. and Schmidt, P. and Slater, J. A. and Trupke, M. and Uhl, K. and Weimann, T. and Wieczorek, W. and Aspelmeyer, M.},
  journal = {Phys. Rev. Lett.},
  volume = {131},
  issue = {4},
  pages = {043603},
  numpages = {7},
  year = {2023},
  month = {Jul},
  publisher = {American Physical Society},
  doi = {10.1103/PhysRevLett.131.043603},
  url = {https://link.aps.org/doi/10.1103/PhysRevLett.131.043603}
}

@article{Melo_2024,
	author = {Melo, Bruno and T. Cuairan, Marc and Tomassi, Gr{\'e}goire F. M. and Meyer, Nadine and Quidant, Romain},
	date = {2024/09/01},
	doi = {10.1038/s41565-024-01677-3},
	id = {Melo2024},
	isbn = {1748-3395},
	journal = {Nature Nanotechnology},
	number = {9},
	pages = {1270--1276},
	title = {Vacuum levitation and motion control on chip},
	url = {https://doi.org/10.1038/s41565-024-01677-3},
	volume = {19},
	year = {2024}}

@article{Bonvin_2024_june,
  title = {State Expansion of a Levitated Nanoparticle in a Dark Harmonic Potential},
  author = {Bonvin, Eric and Devaud, Louisiane and Rossi, Massimiliano and Militaru, Andrei and Dania, Lorenzo and Bykov, Dmitry S. and Romero-Isart, Oriol and Northup, Tracy E. and Novotny, Lukas and Frimmer, Martin},
  journal = {Phys. Rev. Lett.},
  volume = {132},
  issue = {25},
  pages = {253602},
  numpages = {7},
  year = {2024},
  month = {Jun},
  publisher = {American Physical Society},
  doi = {10.1103/PhysRevLett.132.253602},
  url = {https://link.aps.org/doi/10.1103/PhysRevLett.132.253602}
}

@article{Bonvin_2024_nov,
  title = {Hybrid Paul-optical trap with large optical access for levitated optomechanics},
  author = {Bonvin, Eric and Devaud, Louisiane and Rossi, Massimiliano and Militaru, Andrei and Dania, Lorenzo and Bykov, Dmitry S. and Teller, Markus and Northup, Tracy E. and Novotny, Lukas and Frimmer, Martin},
  journal = {Phys. Rev. Res.},
  volume = {6},
  issue = {4},
  pages = {043129},
  numpages = {7},
  year = {2024},
  month = {Nov},
  publisher = {American Physical Society},
  doi = {10.1103/PhysRevResearch.6.043129},
  url = {https://link.aps.org/doi/10.1103/PhysRevResearch.6.043129}
}

@article{Hansen_2025,
      title={Optical Interferometric Readout of a Magnetically Levitated Superconducting Microsphere}, 
      author={J. J. Hansen and S. Minniberger and D. Ilk and P. Asenbaum and G. Higgins and R. G. Povey and P. Schmidt and J. Hofer and R. Claessen and M. Aspelmeyer and M. Trupke},
      year={2025},
        journal={arXiv preprint},
      number={arXiv: 2508.11731},
      archivePrefix={arXiv},
      primaryClass={quant-ph},
      url={https://arxiv.org/abs/2508.11731}
}

@article{Bykov_2019,
  title={Direct loading of nanoparticles under high vacuum into a Paul trap for levitodynamical experiments},
  author={Bykov, Dmitry S and Mestres, Pau and Dania, Lorenzo and Schm{\"o}ger, Lisa and Northup, Tracy E},
  journal={Applied Physics Letters},
url={https://pubs.aip.org/aip/apl/article/115/3/034101/37962/Direct-loading-of-nanoparticles-under-high-vacuum},
  volume={115},
  number={3},
  year={2019},
  publisher={AIP Publishing}
}

@article{Goldwater_2019,
	author = {Goldwater, Daniel and Stickler, Benjamin A and Martinetz, Lukas and Northup, Tracy E and Hornberger, Klaus and Millen, James},
	doi = {10.1088/2058-9565/aaf5f3},
	journal = {Quantum Science and Technology},
	month = {jan},
	number = {2},
	pages = {024003},
	publisher = {IOP Publishing},
	title = {Levitated electromechanics: all-electrical cooling of charged nano- and micro-particles},
	url = {https://dx.doi.org/10.1088/2058-9565/aaf5f3},
	volume = {4},
	year = {2019}}

@inproceedings{Dania_2019,
author = {Lorenzo Dania and Dmitry Bykov and Pau Mestres and Tracy E. Northup},
title = {{Manipulating a charged nanoparticle in a Paul trap for ion-assisted levitated optomechanics}},
volume = {11083},
booktitle = {Optical Trapping and Optical Micromanipulation XVI},
editor = {Kishan Dholakia and Gabriel C. Spalding},
organization = {International Society for Optics and Photonics},
publisher = {SPIE},
pages = {1108335},
year = {2019},
doi = {10.1117/12.2529127},
URL = {https://doi.org/10.1117/12.2529127}
}

@article{Bullier_2020,
	author = {Bullier, N P and Pontin, A and Barker, P F},
	doi = {10.1088/1361-6463/ab71a7},
	journal = {Journal of Physics D: Applied Physics},
	month = {feb},
	number = {17},
	pages = {175302},
	publisher = {IOP Publishing},
	title = {Characterisation of a charged particle levitated nano-oscillator},
	url = {https://dx.doi.org/10.1088/1361-6463/ab71a7},
	volume = {53},
	year = {2020}}

@article{Martinetz_2020,
	author = {Martinetz, Lukas and Hornberger, Klaus and Millen, James and Kim, M. S. and Stickler, Benjamin A.},
	date = {2020/12/11},
	doi = {10.1038/s41534-020-00333-7},
	id = {Martinetz2020},
	isbn = {2056-6387},
	journal = {npj Quantum Information},
	number = {1},
	pages = {101},
	title = {Quantum electromechanics with levitated nanoparticles},
	url = {https://doi.org/10.1038/s41534-020-00333-7},
	volume = {6},
	year = {2020}}

@article{Dania_2021,
  title = {Optical and electrical feedback cooling of a silica nanoparticle levitated in a Paul trap},
  author = {Dania, Lorenzo and Bykov, Dmitry S. and Knoll, Matthias and Mestres, Pau and Northup, Tracy E.},
  journal = {Phys. Rev. Res.},
  volume = {3},
  issue = {1},
  pages = {013018},
  numpages = {12},
  year = {2021},
  month = {Jan},
  publisher = {American Physical Society},
  doi = {10.1103/PhysRevResearch.3.013018},
  url = {https://link.aps.org/doi/10.1103/PhysRevResearch.3.013018}
}

@article{Dania_2024,
  title = {Ultrahigh Quality Factor of a Levitated Nanomechanical Oscillator},
  author = {Dania, Lorenzo and Bykov, Dmitry S. and Goschin, Florian and Teller, Markus and Kassid, Abderrahmane and Northup, Tracy E.},
  journal = {Phys. Rev. Lett.},
  volume = {132},
  issue = {13},
  pages = {133602},
  numpages = {7},
  year = {2024},
  month = {Mar},
  publisher = {American Physical Society},
  doi = {10.1103/PhysRevLett.132.133602},
  url = {https://link.aps.org/doi/10.1103/PhysRevLett.132.133602}
}

@article{Bykov_2025,
  title = {Nanoparticle Stored with an Atomic Ion in a Linear Paul Trap},
  author = {Bykov, Dmitry S. and Dania, Lorenzo and Goschin, Florian and Northup, Tracy E.},
  journal = {Phys. Rev. Lett.},
  volume = {135},
  issue = {21},
  pages = {213602},
  numpages = {6},
  year = {2025},
  month = {Nov},
  publisher = {American Physical Society},
  doi = {10.1103/5lzj-f61x},
  url = {https://link.aps.org/doi/10.1103/5lzj-f61x}
}

@article{Seberson_2020,
	author = {T. Seberson and Peng Ju and Jonghoon Ahn and Jaehoon Bang and Tongcang Li and F. Robicheaux},
	journal = {J. Opt. Soc. Am. B},
	month = {Dec},
	number = {12},
	pages = {3714--3720},
	publisher = {Optica Publishing Group},
	title = {Simulation of sympathetic cooling an optically levitated magnetic nanoparticle via coupling to a cold atomic gas},
	url = {https://opg.optica.org/josab/abstract.cfm?URI=josab-37-12-3714},
	volume = {37},
	year = {2020}}

@article{Penny_2023,
  title = {Sympathetic cooling and squeezing of two colevitated nanoparticles},
  author = {Penny, T. W. and Pontin, A. and Barker, P. F.},
  journal = {Phys. Rev. Res.},
  volume = {5},
  issue = {1},
  pages = {013070},
  numpages = {8},
  year = {2023},
  month = {Jan},
  publisher = {American Physical Society},
  doi = {10.1103/PhysRevResearch.5.013070},
  url = {https://link.aps.org/doi/10.1103/PhysRevResearch.5.013070}
}

@article{Bykov_2023,
	author = {Dmitry S. Bykov and Lorenzo Dania and Florian Goschin and Tracy E. Northup},
	doi = {10.1364/OPTICA.481076},
	journal = {Optica},
	month = {Apr},
	number = {4},
	pages = {438--442},
	publisher = {Optica Publishing Group},
	title = {3D sympathetic cooling and detection of levitated nanoparticles},
	url = {https://opg.optica.org/optica/abstract.cfm?URI=optica-10-4-438},
	volume = {10},
	year = {2023}}

@article{Pflanzer_2013,
  title = {Optomechanics assisted by a qubit: From dissipative state preparation to many-partite systems},
  author = {Pflanzer, Anika C. and Romero-Isart, Oriol and Cirac, J. Ignacio},
  journal = {Phys. Rev. A},
  volume = {88},
  issue = {3},
  pages = {033804},
  numpages = {8},
  year = {2013},
  month = {Sep},
  publisher = {American Physical Society},
  doi = {10.1103/PhysRevA.88.033804},
  url = {https://link.aps.org/doi/10.1103/PhysRevA.88.033804}
}

@article{Toro_2021,
  title = {Creating atom-nanoparticle quantum superpositions},
  author = {Toro\ifmmode \check{s}\else \v{s}\fi{}, M. and Bose, S. and Barker, P. F.},
  journal = {Phys. Rev. Res.},
  volume = {3},
  issue = {3},
  pages = {033218},
  numpages = {11},
  year = {2021},
  month = {Sep},
  publisher = {American Physical Society},
  doi = {10.1103/PhysRevResearch.3.033218},
  url = {https://link.aps.org/doi/10.1103/PhysRevResearch.3.033218}
}

@misc{Schut_2025,
      title={Proposal for macroscopic delocalisation of a large mass in a RF trap}, 
      author={Martine Schut and Valerio Scarani},
      year={2025},
      eprint={2509.17081},
      archivePrefix={arXiv},
      primaryClass={quant-ph},
      url={https://arxiv.org/abs/2509.17081}, 
}

@article{Dehmelt_1995,
    doi = {10.1088/0031-8949/1995/T59/060},
    url = {https://doi.org/10.1088/0031-8949/1995/T59/060},
    year = {1995},
    month = {jan},
    publisher = {},
    volume = {1995},
    number = {T59},
    pages = {423},
    author = {Hans Dehmelt},
    title = {Economic synthesis and precision spectroscopy of anti-molecular hydrogen ions in Paul trap},
    journal = {Physica Scripta}
}

@article{Trypogeorgos_2016,
  title = {Cotrapping different species in ion traps using multiple radio frequencies},
  author = {Trypogeorgos, Dimitris and Foot, Christopher J.},
  journal = {Phys. Rev. A},
  volume = {94},
  issue = {2},
  pages = {023609},
  numpages = {9},
  year = {2016},
  month = {Aug},
  publisher = {American Physical Society},
  doi = {10.1103/PhysRevA.94.023609},
  url = {https://link.aps.org/doi/10.1103/PhysRevA.94.023609}
}

@article{Leefer_2016,
	author = {Leefer, Nathan and Krimmel, Kai and Bertsche, William and Budker, Dmitry and Fajans, Joel and Folman, Ron and H{\"a}ffner, Hartmut and Schmidt-Kaler, Ferdinand},
	date = {2016/12/28},
	doi = {10.1007/s10751-016-1388-0},
	id = {Leefer2016},
	isbn = {1572-9540},
	journal = {Hyperfine Interactions},
	number = {1},
	pages = {12},
	title = {Investigation of two-frequency Paul traps for antihydrogen production},
	url = {https://doi.org/10.1007/s10751-016-1388-0},
	volume = {238},
	year = {2016}}

@article{Amore_2003,
	author = {Paolo Amore and Alfredo Aranda},
	doi = {https://doi.org/10.1016/j.physleta.2003.08.001},
	issn = {0375-9601},
	journal = {Physics Letters A},
	number = {3},
	pages = {218-225},
	title = {Presenting a new method for the solution of nonlinear problems},
	url = {https://www.sciencedirect.com/science/article/pii/S0375960103011794},
	volume = {316},
	year = {2003}
}

@article{Saxena_2018,
  author={Saxena, Varun and Shah, Kushal},
  journal={IEEE Transactions on Plasma Science}, 
  title={Plasma Dynamics in a Dual-Frequency Paul Trap Using Tsallis Distribution}, 
  year={2018},
  volume={46},
  number={3},
  pages={474-481},
  doi={10.1109/TPS.2018.2803799}
}

@article{Leibfried_2003,
  title = {Quantum dynamics of single trapped ions},
  author = {Leibfried, D. and Blatt, R. and Monroe, C. and Wineland, D.},
  journal = {Rev. Mod. Phys.},
  volume = {75},
  issue = {1},
  pages = {281--324},
  numpages = {0},
  year = {2003},
  month = {Mar},
  publisher = {American Physical Society},
  doi = {10.1103/RevModPhys.75.281},
  url = {https://link.aps.org/doi/10.1103/RevModPhys.75.281}
}

@article{Berkeland_1998,
    author = {Berkeland, D. J. and Miller, J. D. and Bergquist, J. C. and Itano, W. M. and Wineland, D. J.},
    title = {Minimization of ion micromotion in a Paul trap},
    journal = {Journal of Applied Physics},
    volume = {83},
    number = {10},
    pages = {5025-5033},
    year = {1998},
    month = {05},
    issn = {0021-8979},
    doi = {10.1063/1.367318},
    url = {https://doi.org/10.1063/1.367318},
}

@book{Ghosh_1995,
  author    = {Pradip K. Ghosh},
  title     = {Ion Traps},
  year      = {1995},
  publisher = {Oxford University Press},
  address   = {Oxford},
  doi       = {10.1093/oso/9780198539957.001.0001},
  isbn      = {9780198539957},
  url       = {https://doi.org/10.1093/oso/9780198539957.001.0001}
}

@article{Lanchares_2014,
  title={On the stability of Hamiltonian dynamical systems},
  author={Lanchares, V{\'\i}ctor},
  journal={Monograf{\'\i}as Matem{\'a}ticas Garc{\'\i}a de Galdeano},
  volume={39},
  url= {https://www.unirioja.es/cu/mheras/MMGG2014.pdf},
  pages={155--166},
  year={2014}
}

@article{Kustura_2019,
  title = {Quadratic quantum Hamiltonians: General canonical transformation to a normal form},
  author = {Kustura, Katja and Rusconi, Cosimo C. and Romero-Isart, Oriol},
  journal = {Phys. Rev. A},
  volume = {99},
  issue = {2},
  pages = {022130},
  numpages = {14},
  year = {2019},
  month = {Feb},
  publisher = {American Physical Society},
  doi = {10.1103/PhysRevA.99.022130},
  url = {https://link.aps.org/doi/10.1103/PhysRevA.99.022130}
}

@article{CGB_2019,
  title = {Theory for cavity cooling of levitated nanoparticles via coherent scattering: Master equation approach},
  author = {Gonzalez-Ballestero, C. and Maurer, P. and Windey, D. and Novotny, L. and Reimann, R. and Romero-Isart, O.},
  journal = {Phys. Rev. A},
  volume = {100},
  issue = {1},
  pages = {013805},
  numpages = {25},
  year = {2019},
  month = {Jul},
  publisher = {American Physical Society},
  doi = {10.1103/PhysRevA.100.013805},
  url = {https://link.aps.org/doi/10.1103/PhysRevA.100.013805}
}

@article{Beresnev_1990, 
    title={Motion of a spherical particle in a rarefied gas. Part 2. Drag and thermal polarization}, 
    volume={219},
    DOI={10.1017/S0022112090003007}, 
    journal={Journal of Fluid Mechanics}, 
    author={Beresnev, S. A. and Chernyak, V. G. and Fomyagin, G. A.}, 
    year={1990},
    pages={405–421}
}

@article{Li_2011,
	author = {Li, Tongcang and Kheifets, Simon and Raizen, Mark G.},
	date = {2011/07/01},
	doi = {10.1038/nphys1952},
	id = {Li2011},
	isbn = {1745-2481},
	journal = {Nature Physics},
	number = {7},
	pages = {527--530},
	title = {Millikelvin cooling of an optically trapped microsphere in vacuum},
	url = {https://doi.org/10.1038/nphys1952},
	volume = {7},
	year = {2011}
}

@article{Eschner_2003,
	author = {J\"{u}rgen Eschner and Giovanna Morigi and Ferdinand Schmidt-Kaler and Rainer Blatt},
	doi = {10.1364/JOSAB.20.001003},
	journal = {J. Opt. Soc. Am. B},
	month = {May},
	number = {5},
	pages = {1003--1015},
	publisher = {Optica Publishing Group},
	title = {Laser cooling of trapped ions},
	url = {https://opg.optica.org/josab/abstract.cfm?URI=josab-20-5-1003},
	volume = {20},
	year = {2003}
}

@article{Diosi_1995,
doi = {10.1209/0295-5075/30/2/001},
url = {https://doi.org/10.1209/0295-5075/30/2/001},
year = {1995},
month = {apr},
publisher = {},
volume = {30},
number = {2},
pages = {63},
author = {L. Diósi},
title = {Quantum Master Equation of a Particle in a Gas Environment},
journal = {Europhysics Letters}
}

@article{MaurerPRA2023,
  title = {Quantum theory of light interaction with a Lorenz-Mie particle: Optical detection and three-dimensional ground-state cooling},
  author = {Maurer, Patrick and Gonzalez-Ballestero, Carlos and Romero-Isart, Oriol},
  journal = {Phys. Rev. A},
  volume = {108},
  issue = {3},
  pages = {033714},
  numpages = {26},
  year = {2023},
  month = {Sep},
  publisher = {American Physical Society},
  doi = {10.1103/PhysRevA.108.033714},
  url = {https://link.aps.org/doi/10.1103/PhysRevA.108.033714}
}

@book{Andreevich_1975,
author="Iakubovich, V. A. and Starzhinskiĭ, V. M.",
title="Linear differential equations with periodic coefficients",
publisher="Wiley and Israel Program for Scientific Translations",
year="1975",
URL="https://cir.nii.ac.jp/crid/1970586434894944166"
}

@article{Slane_2019,
  title={Analysis of periodic nonautonomous inhomogeneous systems},
  author={Slane, Jean and Tragesser, Steven},
  journal={Nonlinear Dynamics and Systems Theory},
  volume={11},
  number={2},
  pages={183--198},
  year={2011},
  url={https://e-ndst.kiev.ua/v11n2/8(35).pdf}
}

@article{Paris_2003,
  title = {Purity of Gaussian states: Measurement schemes and time evolution in noisy channels},
  author = {Paris, Matteo G. A. and Illuminati, Fabrizio and Serafini, Alessio and De Siena, Silvio},
  journal = {Phys. Rev. A},
  volume = {68},
  issue = {1},
  pages = {012314},
  numpages = {9},
  year = {2003},
  month = {Jul},
  publisher = {American Physical Society},
  doi = {10.1103/PhysRevA.68.012314},
  url = {https://link.aps.org/doi/10.1103/PhysRevA.68.012314}
}

@article{Barthel_2022,
	author = {Barthel, Thomas and Zhang, Yikang},
	doi = {10.1088/1742-5468/ac8e5c},
	journal = {Journal of Statistical Mechanics: Theory and Experiment},
	month = {nov},
	number = {11},
	pages = {113101},
	publisher = {IOP Publishing and SISSA},
	title = {Solving quasi-free and quadratic Lindblad master equations for open fermionic and bosonic systems},
	url = {https://doi.org/10.1088/1742-5468/ac8e5c},
	volume = {2022},
	year = {2022}}

@article{Wiseman_1993,
  author  = {Wiseman, H. M. and Milburn, G. J.},
  title   = {Quantum Theory of Optical Feedback via Homodyne Detection},
  journal = {Physical Review Letters},
  volume  = {70},
  number  = {5},
  pages   = {548--551},
  year    = {1993},
  doi     = {10.1103/PhysRevLett.70.548}
}

@article{Cattaneo_2019,
  author  = {Cattaneo, Marco and Giorgi, Gian Luca and Maniscalco, Sabrina and Zambrini, Roberta},
  title   = {Local versus global master equation with common and separate baths: superiority of the global approach in partial secular approximation},
  journal = {New Journal of Physics},
  volume  = {21},
  number  = {11},
  pages   = {113045},
  year    = {2019},
  month   = nov,
  doi     = {10.1088/1367-2630/ab54ac}
}

@article{Tabor2012,
  author  = {Tabor, D. A. and Rajagopal, V. and Lin, Y.-W. and Odom, B.},
  title   = {Suitability of linear quadrupole ion traps for large Coulomb crystals},
  journal = {Applied Physics B},
  volume  = {107},
  pages   = {1097--1104},
  year    = {2012},
  month   = may,
  doi     = {10.1007/s00340-012-5058-x}
}

@article{Deng2015,
  author  = {Deng, K. and Che, H. and Lan, Y. and Ge, Y. P. and Xu, Z. T.
             and Yuan, W. H. and Zhang, J. and Lu, Z. H.},
  title   = {Design of blade-shaped-electrode linear ion traps with reduced
             anharmonic contributions},
  journal = {Journal of Applied Physics},
  volume  = {118},
  number  = {11},
  pages   = {113106},
  year    = {2015},
  doi     = {10.1063/1.4931420}
}

@article{Perdriat2024,
  title = {Rotational Locking of Charged Microparticles in Quadrupole Ion Traps},
  author = {Perdriat, Maxime and Rusconi, Cosimo C. and Delord, Tom and Huillery, Paul and Pellet-Mary, Cl\'ement and Durand, Alrik and Stickler, Benjamin A. and H\'etet, Gabriel},
  journal = {Phys. Rev. Lett.},
  volume = {133},
  issue = {25},
  pages = {253602},
  numpages = {7},
  year = {2024},
  month = {Dec},
  publisher = {American Physical Society},
  doi = {10.1103/PhysRevLett.133.253602},
  url = {https://link.aps.org/doi/10.1103/PhysRevLett.133.253602}
}

@article{vanderLaan_2021,
  author  = {van der Laan, Fons and Tebbenjohanns, Felix and Reimann, Ren\'{e} and Vijayan, Jayadev and Novotny, Lukas and Frimmer, Martin},
  title   = {Sub-{K}elvin Feedback Cooling and Heating Dynamics of an Optically Levitated Librator},
  journal = {Phys. Rev. Lett.},
  volume  = {127},
  pages   = {123605},
  year    = {2021},
  doi     = {10.1103/PhysRevLett.127.123605}
}

@phdthesis{vanderLaan_2022,
  author = {van der Laan, Fons},
  title  = {Rotational Levitodynamics},
  school = {ETH Zurich},
  year   = {2022},
  doi    = {10.3929/ethz-b-000591484}
}

@article{Martinetz_2018,
  author  = {Martinetz, Lukas and Hornberger, Klaus and Stickler, Benjamin A.},
  title   = {Gas-induced friction and diffusion of rigid rotors},
  journal = {Phys. Rev. E},
  volume  = {97},
  pages   = {052112},
  year    = {2018},
  doi     = {10.1103/PhysRevE.97.052112}
}

@article{Stickler_2021,
  author  = {Stickler, Benjamin A. and Hornberger, Klaus and Kim, M. S.},
  title   = {Quantum rotations of nanoparticles},
  journal = {Nature Reviews Physics},
  volume  = {3},
  pages   = {589--597},
  year    = {2021},
  doi     = {10.1038/s42254-021-00335-0}
}

@article{Zielinska_2024,
  title = {Long-Axis Spinning of an Optically Levitated Particle: A Levitated Spinning Top},
  author = {Zieli\ifmmode \acute{n}\else \'{n}\fi{}ska, J. A. and van der Laan, F. and Norrman, A. and Reimann, R. and Frimmer, M. and Novotny, L.},
  journal = {Phys. Rev. Lett.},
  volume = {132},
  issue = {25},
  pages = {253601},
  numpages = {6},
  year = {2024},
  month = {Jun},
  publisher = {American Physical Society},
  doi = {10.1103/PhysRevLett.132.253601},
  url = {https://link.aps.org/doi/10.1103/PhysRevLett.132.253601}
}

\end{document}